\documentclass[12pt,a4paper]{article}
\usepackage{jheppub}
\usepackage{amsmath,amssymb,amsfonts,pstricks,mathtools,setspace}
\usepackage{float}
\usepackage{scalefnt}
\usepackage{wrapfig}
\usepackage{fancybox}
\newcommand{\bea}{\begin{eqnarray}\displaystyle}
\newcommand{\eea}{\end{eqnarray}}
\newcommand{\nn}{\nonumber}
\newcommand{\beq}{\begin{equation}}
\newcommand{\eeq}{\end{equation}}

\newcommand{\figref}[1]{Fig.~\protect\ref{#1}}
{\setlength{\fboxsep}{15pt}
\setlength{\mylength}{\linewidth}%
\addtolength{\mylength}{-2\fboxsep}%
\addtolength{\mylength}{-2\fboxrule}%
\Sbox
\minipage{\mylength}%
\setlength{\abovedisplayskip}{0pt}%
\setlength{\belowdisplayskip}{0pt}%
\equation}%
{\endequation\endminipage\endSbox
\[\fbox{\TheSbox}\]}
\scalefont{2}
\begin{document}

\title{BPS Degeneracies and Superconformal Index
in Diverse Dimensions}
\author[1]{Amer Iqbal,}
\author[2]{Cumrun Vafa}
\affiliation[1]{Department of Physics, LUMS School of Science \& Engineering, U-Block, D.H.A, Lahore, Pakistan.}
\affiliation[1]{Department of Mathematics, LUMS School of Science \& Engineering, U-Block, D.H.A, Lahore, Pakistan.}
\affiliation[2]{Jefferson Physical Laboratory, Harvard University, Cambridge, MA 02138, USA.}
\emailAdd{amer.iqbal@lums.edu.pk}
\emailAdd{vafa@physics.harvard.edu}

\abstract{We present a unifying theme relating BPS partition functions and superconformal indices. In the case with complex
SUSY central
charges (as in ${\cal N}=2$ in $d=4$ and ${\cal N}=(2,2)$ in $d=2$)  the known
results can be reinterpreted as the statement that the BPS partition functions can be used to compute
a specialization of the
superconformal indices.  We argue that in the case with real central charge in the supersymmetry
algebra, as in ${\cal N}=1$ in $d=5$ (or the ${\cal N}=2$ in $d=3$) the BPS
degeneracy
captures the full superconformal index.  Furthermore,
we argue that refined topological strings, which captures 5d BPS
degeneracies of M-theory on CY 3-folds, can be used to compute
5d supersymmetric index including in the sectors with 3d defects
for a large class of 5d superconformal theories.
Moreover,  we provide evidence  that distinct Calabi-Yau
singularities which are expected to lead to the same SCFT yield the same index.}
\maketitle


\section{Introduction}
\scalefont{1.05}
Supersymmetric BPS states have played an important role in many aspects of string theory.  Their
mass is typically protected by SUSY and provides a tool to analyze various limits of string theory.
On the other hand superconformal theories have also figured prominently in many developments
of string theory.  As we deform conformal theory away from
the conformal point, BPS states arise in the resulting theory.  It is natural to ask what is the relation between BPS states that appear and
the properties of the superconformal theory they come from.
In fact there is evidence that the BPS spectrum away from the conformal point
is faithful, and the superconformal theories are entirely captured
by the BPS spectrum.  In particular, we do not have  a single example
of two distinct superconformal theories which give the same BPS spectrum upon deformation.
Of course, not arbitrary BPS spectrum gives
rise to a  consistent theory, and consistency conditions on what the allowed BPS states can be,
has been proposed as a way to classify conformal theories
for ${\cal N}=(2,2)$ in $d=2$ \cite{cecotti1} and ${\cal N}=2$ in $d=4$ \cite{cecotti3,cecotti4} .
If this is the case, it should be possible to recover all the data at the conformal fixed point solely from the
BPS data.
In particular it is natural to ask if the superconformal partition functions such as supersymmetric
indices \cite{Kinney,dolan,shiraz}
are reproducible from the BPS spectrum.

The most natural idea would be to treat BPS states as if they are the elementary
building blocks of the theory and use them to compute the superconformal partition functions.
However the story is not always so simple.  For example for theories
with complex central charge, the BPS spectrum has different chambers
separated by walls.  Nevertheless, as we will review (and partially reinterpret), it is known
 that at least in the cases of
 $d=2$ with
${\cal N}=(2,2)$ \cite{cecotti3} and $d=4$ with ${\cal N}=2$ \cite{vafa3} a specialization of the superconformal index
can be recovered from BPS spectrum in any chamber.

We will provide evidence that the situation is similar but  more powerful in the case of
theories in $d=3,5$ dimensions with Coulomb branch,
with ${\cal N}=2,1$ supersymmetries respectively.  Both of these
cases involve a real central charge.  In these cases we propose that one can recover
the full superconformal index solely from the BPS data in a Coulomb branch of the theory.
In the case of $d=3$ we reinterpret the computations already done as computing
contributions from BPS states.  The main new case involves the superconformal
index in $d=5$.

The basic class of examples we consider is obtained from M-theory on Calabi-Yau threefolds
leading to ${\cal N}=2$ theories in $d=5$ dimensions.   It is known that for these cases the
topological string captures the BPS degeneracies  (corresponding
to M2 branes wrapping 2-cycles) \cite{GV,IHV}.
In addition one can introduce M5 branes wrapping Lagrangian submanifolds
of Calabi-Yau.  These lead to 3 dimensional defects in the 5d theory.  Furthermore it
is known that open topological strings captures the open BPS state degeneracy
for these sectors corresponding to M2 branes ending on M5 branes \cite{OV,GSV}.  We will argue that superconformal
index, i.e. the partition function on $S^1\times S^4$ where the 3d defects
 wrap
$S^1\times S^2$ where $S^2\subset S^4$, can be computed simply by
considering
\bea\nn
\int {dQ_i\over Q_i} {dU_j\over U_j} {\big|} Z_{top}(Q_i,U_j,{\tilde Q_k};q_1,q_2) {\big|}^2\,,
\eea
where $Z_{top}$ is the refined open and closed topological string amplitudes, $Q_i$ correspond to the Wilson line associated with nomalizable K\"ahler moduli of Calabi-Yau,  ${\tilde Q}_k$ is the non-normalizable
K\"ahler moduli, which correspond to mass parameters, $U_j$ correspond to the Wilson
lines for the Lagrangian branes and $(q_{1},q_{2})$ are the two coupling constants
of the refined topological string.   Here complex conjugation sends\footnote{As
we will discuss later, for the defect sector we can turn on monopole
flux which would correspond to allowing $U_j$ to be complex.} $(Q_i,U_j,{\tilde Q}_k; q_1,q_2)\rightarrow
(Q_i^{-1},U_j^{-1},{\tilde Q_k}^{-1};q_1^{-1},q_2^{-1})$.  Furthermore this computation can be
viewed as computing the scattering amplitudes of a string theory
in 4 dimensions proposed recently \cite{vafa}.

A unifying theme seems to emerge about the connection of BPS states to the index, which
can be summarized roughly as follows:  We order the BPS states according to the
phase of their BPS central charge.  In the case of real central charge this simply means dividing the BPS states
to CTP conjugate pairs where one half of the states are on right and the other on the left
of the real line.  In the case of complex central charge this means organizing the states on a circle
according to the phase of the central charge where CTP conjugate pairs are diametrically opposite.
Whether it is real or complex central charge we can consider a `partition function'
of the BPS states where each
BPS state $i$ is represented by an
operator  $\Phi_i$ and we take the product over all the BPS states.  The operator
acts on a different Hilbert space depending on the dimension and the theory in question:
In the 2d case it involves the space of massive vacua,
in the 3d and 5d cases it is the space of flat connections on $S^1$ for the corresponding abelian
gauge groups, and in the 4d case it is the Hilbert space of a $U(1)$  Chern-Simons theory on the Seiberg-Witten curve.

In the complex
central charge case $\Phi_i$ do not commute and we have to order them according
to the phase of the central charge in the SUSY algebra.  In the real central charge case they commute.  Moreover knowing
the contribution for half the states is sufficient, because the CTP conjugate case can be
obtained from them.  Let
\bea\nn
S=\prod_i \Phi_i
\eea
denote the (ordered) product over the BPS states whose phase is on one side.
The full partition over BPS states will take the form
\bea
M=SS^{-t}.
\eea
Then the statement is that
\bea
{\rm Tr}\ M=Z (S^1\times S^{d-1})
\eea
for suitably defined partition function $Z$ of the theory on $S^{1}\times S^{d-1}$ .   For $d=3,5$ this gives the full index and for $d=2,4$ this
gives a specialization of the index.

The intuitive idea for why such a picture holds may be that we can view
operators at the conformal fixed point as being made of the composite
of operators which create BPS states.  In some cases where there is a weak
coupling description of the theory, as in $d=3$ gauge theories, this picture
can be fully justified.

The fact that we propose that the superconformal index in 5 dimensions
can be computed only from the knowledge of BPS particles is surprising
in the following sense:  These theories also have BPS strings.  If we go to the conformal point, we will have
a system of interacting massless particles and tensionless strings.  Upon
going to the Coulomb branch the particles pick up mass and tensionless strings pick up tension.  Moreover the mass
scale for both the interacting strings and the particles are the same \cite{witten}.
What is surprising is that nevertheless the knowledge of only BPS particles is enough
to recapture the full superconformal index in 5 dimensions.  Perhaps this can
be explained by the fact that $S^1\times S^4$ has no 2-cycles for the worldsheet
of BPS strings to wrap around and the properties
of the BPS strings are secretly encoded by the particle states, as far as the index
is concerned.

The organization of this paper is as follows. In section 2 we discuss the relation between superconformal indices and BPS states in two and the four dimensional theories with complex central charges. In section 3 we discuss the three and five dimensional gauge theories with real cental charges, superconformal indices and their relation with BPS states including coupling to the 3d defects.  In section 4 we review the refinement of topological strings and how the refined amplitudes can be calculated. In section 6 we give some examples of index computations for certain 5D theories coming from local CY threefolds including in the presence of 3d defects. In section 7 we present our conclusions.

\section{BPS states and theories with complex central charge in $d=2,4$}
In this section we review (and partially reinterpret) what is known for the relation between BPS states
and superconformal partition functions in the case of ${\cal N}=(2,2)$ theories
in $d=2$ and ${\cal N}=2$ theories in $d=4$.

\subsection{$(2,2)$ theories in $d=2$}
Consider an ${\cal N}=(2,2)$ conformal theory in $d=2$.  In this context we can define
a superconformal index (which is an elliptic genus) \cite{witten2} given by
the following trace in the Ramond sector:
\bea\nn
Z(q,z)={\rm Tr}(-1)^F z^{J_L}q^{H_L}{\overline q}^{H_R}
\eea
where $J_L$ is the left-moving $U(1)_R$ charge and $F=F_{L}-F_{R}$, $F_{L,R}$ being the fermion numbers of the left and the right movers.  Since the Ramond sector is supersymmetric, by SUSY argument as in the Witten index,
the above index only depends on $q,z$ and is independent of moduli of conformal theory.
It receives contributions from all the states which are ground states
of the $H_R$ and it is an arbitrary eigenstate of $H_L$.   Note that in the limit
$q\rightarrow 0$ this receives contribution only from the ground state $H_L=H_R=0$.
In this case $Z(0,z) $ simply computes the partition function of the ground states
in the Ramond sector weighted by their R-charge $J_L$.

We will consider a subset of ${\cal N}=(2,2)$ theories which admit deformations
which flow in the IR to a trivial theory.    For this to be possible in particular $J_L-J_R\in {\bf Z}$ and
the ground states have equal $J_L,J_R$ charges.  The index of such theories,
which is also equal to the number of distinct vacua upon mass deformations is $N=Z(0,1)$.

For special values of $z$,  the index simplifies and becomes $q$-independent:  Let
\bea\nn
z={\rm exp}(2\pi ik)
\eea
note that since $J_L$ is not necessarily an integer, putting $z={\rm exp}(2\pi ik)$ is not
the same as $z=1$.   Moreover in this limit the left-moving supercharges also
commute with the elements in the trace and the partition function is $q$ independent,
and in particular can be evaluated by taking the $q\rightarrow 0$ limit:
\bea\nn
Z(q,{\rm exp}(2\pi i k))=Z(0,{\rm exp}(2\pi i k))=Z_k
\eea
In particular as shown in \cite{vafa2} using spectral flow, $Z_k$ counts the index of the theory relative to $(G^+_k,{\overline G}^+_0)$, where $(G,{\overline G})$ refer to (left,right)-moving supercharges
in the Ramond sector.

This theory will have BPS kinks connecting the various vacua.
The number
of kinks depends on how we deform the superconformal theory to the
massive ones, and there are domain walls in parameter space where
the BPS degeneracies change  \cite{ken}.
  Let $m_{ij}$
be the number of kinks connecting the $i$-th vacuum to the $j$-th one, taking
into account the $(-1)^F$ acting on the lowest state of the multiplet.  BPS
kinks come with complex central charges.  Order the vacua such that
the phase of the corresponding central charges $Z_{i,i+1}$ goes
counter-clockwise as we increase $i$.  In this basis
let $A$ be the upper triangular matrix given by $A_{ij}=m_{ij}$ for
each $i<j$.  Consider the matrix
\bea\nn
S=1-A
\eea
and furthermore construct the matrix
\beq\label{mono}
M=S S^{-t}=(1-A)\cdot {1\over {1-A^t}}
\eeq
where $S^{-t}$ is the inverse transpose of $S$.  Since $A$ is upper triangular we have
\bea\nn
S^{-t}=1+A^t+A^{2t}+...+A^{(N-1)t}\,.
\eea
Wall crossing formula for the BPS states \cite{ken} imply that the eigenvalues
of $M$ do not depend on which chamber we compute it in
(even though $S$ does change).   So it is purely a property of the
conformal fixed point.  Moreover, using $tt^*$ equations \cite{cecotti2} it was shown in \cite{cecotti1} that\footnote{Furthermore it was shown how this can be refined to compute the $Z(0,z)$ for
arbitrary $z$. }
\bea
{\rm Tr} M^k&=&{\rm Tr}_{H=0} \ {\rm exp}(2\pi i k J_L)\\\nn
&=& Z_k\,.
\eea
 Moreover this was used as a starting point of a program to classify
${\cal N}=(2,2)$ theories in $d=2$. For a recent discussion of the meaning of this
relation see \cite{cecotti5}.

\hskip-1cm \textcolor[rgb]{1.00,0.50,0.25}{\rule{6.45in}{0.02in}}\
{\bf An Example}: As an example consider the case of LG theory with superpotential $W=\tfrac{1}{3}x^3$ for which a conformal fixed point is expected \cite{emil,nick}. The spectrum of the R-charges at the conformal point is $\pm \frac{1}{6}$. The chiral ring consists of $\{1,x\}$ and when the theory is deformed so that the superpotential becomes $W=\tfrac{1}{3}x^3-a\,x$ we get two vacua for $x_{\pm}=\pm \sqrt{a}$. There is a single BPS kink connecting them therefore
\bea\nn
S=\left(
    \begin{array}{cc}
      1 & -1 \\
      0 & \,1 \\
    \end{array}
  \right)\,.
\eea

\begin{figure}[h]
  \centering
  \includegraphics[width=1.7in]{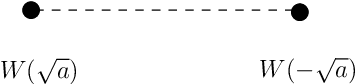}\\
\end{figure}
  
$M=SS^{-t}$ has two eigenvalues $\mbox{exp}(\pm \frac{2\pi i}{6})$ which
agrees with the spectrum of the R-charges of the theory at the conformal point.

\hskip-1cm \textcolor[rgb]{1.00,0.50,0.25}{\rule{6.45in}{0.02in}}\

It is interesting to note that Eq.(\ref{mono}) has the structure of the partition function
of fermions and bosons.  It is as if we are constructing composite operators
from the fields creating the kinks.  Moreover
consider the kink operators
placed on a circle ordered by the phase of their central charge and the
ones on the left semi-circle are fermionic and the ones on the right-half
are bosonic.  Then the ${\rm Tr} M$  can be viewed as the totality
of operators we can make out of them which can be placed on a circle
(i.e. start from one vacuum and end on the same vacuum).  This structure
will repeat, as we shall see in all the other dimensions where we connect
BPS degeneracies with partition functions at superconformal points.

\subsection{BPS states and ${\cal N}=2$, in $d=4$ dimensions}

The connection between degeneracies of BPS states for ${\cal N}=2$ theories
in $d=4$ and certain partition functions at the superconformal point was
found in \cite{vafa3}.  We consider the
theory in the background
involving $S^1\times MC_q$ where $MC_q$ is the Melvin cigar:
$MC_q$ is given by ${\bf C}\times S^1$ where we rotate ${\bf C}$ by $q$ as we go around $S^1$.
Moreover as we go around the other $S^1$ we twist by $t^{r-R}$ where $r$ is the
extra r-charge which is a symmetry at the conformal point and $R$ is a Cartan
in the $SU(2)_R$.    The $MC_q$ can be viewed
topologically as ${1\over 2} S^3$ with squashing parameter $q$.  We will denote this
by
\bea\nn
MC_q={1\over 2}S^3_q\,.
\eea
One considers the partition function on $S^1\times {1\over 2} S^3_q$ which can
be represented in the operator formulation as (suppressing the irrelevant $e^{-\beta H}$)
\bea
Z(t,q)={\rm Tr}_{{1\over 2}S^3_q}(-1)^F t^{r-R}\,.
\eea
We now explain the relation of this partition function with the deformed theory.   Each BPS state is characterized by a charge $\gamma$ which belongs
to the lattice of electric and magnetic charges.  Note that this lattice
has a canonical skew-symmetric product pairing the electric with
the corresponding magnetic charges.  Consider the quantum torus
algebra given by introducing for each element $\gamma$ of the lattice
an operator $U_\gamma$ satisfying\footnote{When the ${\cal N}=2$, $d=4$ theory is realized in terms of an M5-brane wrapping $\Sigma\times S^{1}$ inside a CY3fold, $S^{1}$ being the time direction, then BPS states are given by M2-branes bounding $\gamma\in H_{1}(\Sigma,\mathbb{Z})$ and $U_{\gamma}$ is the holonomy of the gauge field coming from the B-field on the M5-brane reduced along the cycles of $\Sigma$ \cite{vafa3}.}
\bea\nn
U_\gamma U_\beta =q^{\langle \gamma, \beta \rangle} U_\beta U_\gamma\,.
\eea
For each BPS state of charge $\gamma$ and spin $s$ introduce the operator
\bea
\Phi (\gamma,s)=\prod_n (1-q^{n+s+{1\over 2}} U_\gamma)^{(-1)^{2s}}\,.
\eea
Consider BPS states whose central charges lie on the upper half-plane.
\bea
S=T\big( \prod_{BPS-upper} \Phi (\gamma_i,s_i) \big) \,,
\eea
where $T$ denote ordering the product in the order of the phases of the central
charges as it goes in a counter-clockwise direction.
Furthermore consider the matrix
\bea\nn
M=S S^{-t}\,,
\eea
as in the 2d case, where the inverting of $S$ means replacing $U_\gamma\rightarrow
U_{\gamma}^{-1}$ and $q\rightarrow q^{-1}$, $s\rightarrow -s$ and taking the inverse of the products.
Furthermore transposition means the order in the product continues in the order of
increasing phase of central charge.  It was found in \cite{vafa3} that
\bea
{\rm Tr} M^k= Z(t=e^{2\pi ik},q) ={\rm Tr}_{{1\over 2}S^3_q}(-1)^F e^{2\pi i kr}\,.
\eea
The fact that this gives the same result in all chambers follows from the work
of Kontsevich-Soibelman \cite{KS} and its refinement \cite{soib}.
The similarity of the setup to the 2d case is striking and was explained in \cite{vafa3}.
For alternative derivation see \cite{greg}.

It is tempting to connect this to more standard superconformal index.  In fact
as noted in \cite{vafa3,CCV,sara,DGG} if we consider the double space $S^3_q$, the
partition function on this space gets related to a doubled version of
BPS contributions given by
$${\hat \Phi} (\gamma,s)={\prod_n (1-q^{n+s+{1\over 2}} U_\gamma)^{(-1)^{2s}}\over
\prod_n (1-{\hat q}^{n+s+{1\over 2}} {\hat U}_\gamma)^{(-1)^{2s}}}$$
where $\hat q=exp(-2\pi i/\tau)$ with the parameterizations $q=exp(2\pi i\tau)$,
and ${\hat U}=U^{1\over \tau}$.  It can be checked that ${\hat U}_\gamma$ satisfy
$${\hat U}_\gamma {\hat U}_\beta =q^{\langle \gamma, \beta \rangle} {\hat U}_\beta {\hat U}_\gamma.$$
Moreover $U_\gamma$ and ${\hat U}_\beta$ commute.
Then, it was proposed in \cite{vafa3} that if we consider
$${\hat M}={\hat S}{\hat S}^{-t}$$
where ${\hat S}$ is
constructed out of $\hat \Phi$, then
$${\rm Tr} {\hat M}^k ={\rm Tr}_{S^3_q} (-1)^Fe^{2\pi i kr}$$
It is natural to compare this with the usual superconformal index.  Given the
relation between superconformal index in 4d and the partition function on squashed
$S^3$ \cite{Imamura,Gadde, Dolan}, it is natural to propose\footnote{The combination $J_{12}-J_{34}$ was suggested by the relation with topological strings. However, the correct relation was recently found in \cite{Cecotti:2015lab} and is given by ${\rm Tr}{\hat M}^k={\rm Tr}(-1)^F e^{2\pi i k (r-R)} q^{J_{12}-R}e^{2\pi i(J_{34}-R)}$, which is $t\mapsto e^{2\pi i k}$ and $p\mapsto e^{2\pi i}$ limit of the superconformal index.}
$${\rm Tr}{\hat M}^k={\rm Tr}(-1)^F e^{2\pi i k (r-R)} q^{J_{12}-J_{34}}$$
which can be viewed as a special limit of the $N=2$ superconformal index:
$${\rm Tr}(-1)^F t^{r-R}q^{J_{12}-R}p^{J_{34}-R}$$
with the specialization $pq=1, t=e^{2\pi ik}$.

\section{BPS states and gauge theories with real central charges in $d=3,5$}

In this section we review the computations done for the superconformal
index for gauge theories with ${\cal N}=2$ in $d=3$ and ${\cal N}=1$ in $d=5$.  In both cases
we argue that they can be written entirely in terms of BPS states of the corresponding
theories in the Coulomb branches.  This reinterpretation leads to our general proposal
for relation between BPS states and the index for all superconformal theories in $d=3,5$.

\subsection{Superconformal index in $d=3$, {\cal N}=2}

Here we review the basic statement for computation of superconformal
index for gauge theories on $S^1 \times S^2$  \cite{Kim,Imamura2,Kapustin,DGG}.

Consider a 3d theory with gauge group $G$, and some matter representations ${\cal R}$.
Moreover, depending on what interactions are turned on, certain flavor symmetries
can be introduced.  The superconformal index can be viewed as computation of
$$I_3={\rm Tr}(-1)^F q^{R-J}\textstyle{\prod_{i}} z_i^{F_i}$$
where $J$ is the rotation generator on $S^2$, $R$ denotes the R-charge and
$F_i$ are some flavor charges.  The basic statement is that we can compute $I_3$ simply by taking the contribution of all the fields in the UV to the index, where it can be taken
to be a weakly coupled theory.  Since the index does not change upon
flow, this would give the superconformal index at the conformal point as well.
If we have gauge group factors we can turn on flat connections on $S^1$, which
we denote by $U_i$, which need to be integrated over.  This is equivalent
to projecting to gauge invariant fields.  Moreover, for each flavor
charge we introduce a fugacity $z_i$ around the circle.

The contribution for each particle splits up formally to a square due to CTP
structure of each multiplet.  Let $\Phi_a (z_i,U_j,q) $ be the contribution of one of the particles.
Let the spin of the particle be $s$, and charges $f_i$ under the flavor symmetries,
and charge $p_i$ under the gauge symmetries.   Then\footnote{Here we are turning
off the fugacity associated with monopole number which can be viewed as complexification
of $U_j$ \cite{DGG}.}
$$\Phi_a(z_i,U_j,q)=\prod_n {(1-q^{n+\delta_a+{1\over 2} }U_j^{p_j} z_i^{f_i})}^{(-1)^{2s}}$$
where $\delta_a$ is the $R$-charge of the field (and for free theory
gets identified with $s$).
Including the CTP conjugate is the same as introducing  $\Phi_a^{-t}=1/\Phi_a(z_i^{-1},U_j^{-1},q)$.
Let
$$S=\prod_a \Phi_a$$
Then the index can be written as\footnote{The integration is over the Cartan of $U(1)^n$.  This
is also true in the non-abelian case where the extra measure factors can be viewed as arising
from the contributions $\Phi_a$ of the massive gauge particles of the non-abelian group
in the Coulomb branch.}
$$I_3={\rm Tr}\ M={\rm Tr} \ SS^{-t}=\int {dU_j \over U_j} \prod_a \Phi_a(z_i,U_j,q)\cdot  {1\over \Phi_a(z_i^{-1},U_j^{-1},q)}$$
which has the same structure as what we had seen in $d=2,4$.
Indeed if we go to the Coulomb branch the basic field become BPS states
and so this can also be viewed as computation using the BPS states\footnote{More precisely
what we mean by this is that if we ungauge the $U(1)$'s, the BPS partition function
of the global symmetries determine what are the BPS states.  The index for the $U(1)$'s
which we gauge is determined entirely in terms of them.}, in the
same sense as we had seen in $d=2,4$.
Note that at least formally this can be written in the form
$$I_3=\int {dU_j \over U_j} {\bigg |}\prod_a \Phi_a(z_i,U_j,q) {\bigg |}^2$$
using the fact that (not worrying about regions of convergence of $q$)
$$\Phi(z_i^{-1},U_j^{-1},q^{-1})={1\over \Phi(z_i^{-1},U_j^{-1},q)}$$
This computes the index at zero monopole number.
One can also include the effect of the global symmetries associated with shifting
the dual photon.  This can be done most naturally by considering a generalized index \cite{Kapustin}
with fixed monopole numbers $m_j$.  This can be shown to be equivalent  \cite{DGG} to viewing
holonomies as complex, shifting $U_j\rightarrow U_jX_j$
where $X_j$ is viewed as real and at the end, after taking $|...|^2$ substituted by $X_j=q^{m_j/2}$.

\subsection{${\cal N}=1$, $d=5	$ and BPS states}

The superconformal index in $d=5$ is defined \cite{shiraz} by the twisted partition function on $S^1\times S^4$:
$$I_5={\rm Tr} (-1)^F q_1^{J_{12}-R}q_2^{J_{34}-R}z_j^{f_j}$$
where $J_{12}$ and $J_{34}$ are the rotations of two planes in $S^4$ and $R$ denotes
the Cartan of the $SU(2)$ R-symmetry, and $f_i$ denote flavor symmetries.
The fact that there are non-trivial ${\cal N}=1$ superconformal theories
has been argued from many different viewpoints \cite{Seiberg,DKV,MS,witten,MIS}.
There are non-trivial superconformal field theories whose existence is signaled
by the existence of massless particles and tensionless strings.
Moreover, as argued in \cite{Seiberg} many superconformal theories deform
upon mass deformations to gauge theories.  In turn, in the IR limit the gauge theories
become weakly coupled, and one can use this weakly coupled IR theory to compute
the index.  Since the index is independent of deformations
this can be used to recover the index at the conformal point.  This idea has
been considered in \cite{KKL} where the superconformal index
for some theories were computed using localization techniques.  This includes that of $SU(2)$ with up to $N_f=7$ fundamental matter.  Moreover the expected $E_{N_f+1}$ symmetry
of these theories was successfully tested.  The basic structure of the answer can be recast,
which we discuss in more detail in section 6, as
$$I_5=\int {dU_i\over U_i} \big|Z_{5d}^{Nekrasov}(U_i,z_j;q_1,q_2)\big|^2$$
where $Z_{5d}^{Nekrasov} $ denotes the Nekrasov partition function
for the 4d theories coming from compactification of the theory on $S^1$,
and $U_i$ denote the holonomy of the gauge group along $S^1$,  and $z_j$ are
exponential of mass parameters and the instanton number (which is one of the flavor
symmetries).  Moreover in the above formula the $|...|^2$ involves complex
conjugating the $U_i,z_j\rightarrow U_i^{-1},z_j^{-1}$ but keeping $q_{1,2}$ unchanged.
 Of course this result was already anticipated by the computation
of Pestun \cite{pestun} relating 4d Nekrasov partition function with gauge theory
partition function on $S^4$.  This can be viewed as a special instance
of that general argument where the argument is applied to the 4d theory
obtained by compactification from 5d.

The question is what is the relation of this index with BPS states?  Unlike
the 3d case, where the basic fields can be viewed as BPS states in the Coulomb branch,
in the 5d case the gauge fields and matter fields are not the only BPS states.
Indeed this is consistent with the fact that $I_5$ is considerably more complicated
than the $3d$ case where the index is given by treating the basic
fields as the only relevant ingredients for the computations.  Indeed
there are infinitely many BPS states in this case.  The question is
whether $I_5$ can be reinterpreted just in terms of BPS states, as was
the case in $d=2,3,4$?

As is well known the partition function
of refined topological strings on a CY which engineers the corresponding
gauge theory \cite{KMV} is identical with Nekrasov's partition function.  Therefore
we can interpret the above statement as
$$I_5=\int {dU_i\over U_i} \big|Z_{CY}^{top.}(U_i,z_j;q_1,q_2)\big|^2$$
On the other hand, it is known that topological strings captures BPS degeneracies \cite{GV}
(see \cite{IHV} for the refined version):
$$Z^{top}=\prod_{s_i,n_i,m_j}\prod_{m,n=1}^{\infty} (1-q_1^{m+s_1+{1\over 2}}q_2^{n+s_2+{1\over 2} }U_i^{n_i}z_j^{m_j})^{(-1)^{2s} N_{s_1,s_2,n_i,m_j}}$$
where $N_{s_1,s_2,n_i,m_j}$ is the BPS degeneracy with $SO(4)$ spins $s_i$
written in an orthogonal basis of Cartan, gauge charges
$n_i$ and flavor charge $m_j$ (where in topological string $(n_i,m_j)$ translate
to an element of $H_2$ of CY where the M2 brane wraps to give rise to BPS state).
Thus we can view this $Z^{top}$ as a partition function of BPS particles:
$$Z^{top}=\prod_{i\in BPS} \Phi_i =S$$
with $\Phi$ identified as the above, counting the BPS states as if they are the elementary
building blocks of the theory, even though there is no weak coupling Lagrangian which
describes them as fundamental fields.  Nevertheless they seem to behave as such.
Moreover $S^{-t}$ is given by
$$S^{-t}= Z^{top}(q_1^{-1},q_2^{-1},U_i^{-1},z_j^{-1})={1\over Z^{top}(q_1,q_2^{-1},U_i^{-1},z_j^{-1})}=
Z^{top}(q_1,q_2,U_i^{-1},z_j^{-1})$$

The proof of this is given in section 4 above Eq.(\ref{sym3}) when we discuss the properties of the refined partition function.

Therefore we can again write the index as
$$I_5={\rm Tr}\ M=Tr\ S S^{-t}=\int {dU_i\over U_i}|Z^{top}(q_1,q_2,U_i,z_j)|^2$$
Thus we have a unified picture in $d=2,3,4,5$ on the relation between BPS
states and supersymmetric partition functions.

\subsection{Inclusion of codimension 2 defects}

In the context of topological strings we can also consider M5 branes wrapping
special Lagrangian submanifolds.  These correspond to 3d defects in gauge theory,
giving the analog of surface operators in the context of 4d gauge theory \cite{GW1,GW2}.
We will describe the detailed definition of them shortly.    We can then
ask how one may compute the index of the 5d theory in the presence of 3d defects.
This fits nicely with the above formalism by simply combining the degrees of freedom of the BPS states involving M2 branes
ending on M5 branes, which open topological string counts \cite{OV,GSV}:
$$I_{5,3}=\int {dU_i\over U_i}{dV_j\over V_j}\big| Z^{top}_{open,closed}(q_1,q_2,U_i,V_j,z_k)\big|^2$$
where $U_i,V_j$ are the bulk and defect holonomies around $S^1$ respectively and
$z_k$ are the flavor symmetries.
$z_k$ correspond to K\"ahler classes in the Calabi-Yau. This computes
the index in the zero monopole number sector.
 To obtain the generalized
index of \cite{Kapustin,DGG} with fixed monopole numbers $m_j$, it suffices to take $V_j$ to have
a real piece $V_j\rightarrow V_j X_j$ and substituting, {\it after} taking the $|...|^2$,
$X_j=q_1^{m_j/2}$, where we have taken the M5-brane to be in the 12-plane.

\begin{figure}[h]
  \centering
  \includegraphics[width=3.5in]{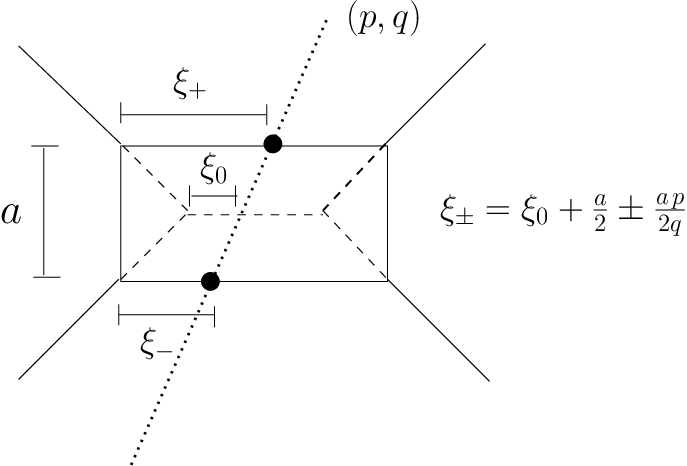}\\
  \caption{The geometry of Lagrangian brane on local $\mathbb{P}^{1}\times \mathbb{P}^{1}$.
    Here we have chosen the spectator brane to be $(p,q)$ with slope $p/q$.
The CS level on the brane is at $k=q$.
  The Lagrangian brane is suspended from the spectator brane at either of the
  two points (denoted by black dots).
  The Coulomb branch parameter is labeled by $a$.
  Moreover the slope being $p/q$ affects how the
  effective FI terms $\xi_{\pm}=\xi_0+{a\over 2}\pm {ap\over 2q}$ change with $a$. }\label{lagg1}
\end{figure}

Next we discuss in more detail
the connection between M5 branes wrapping Lagrangian submanifolds and gauge theoretic defects (see also \cite{hol}).
M5 branes wrapped on special Lagrangian submanifolds and filling an $\mathbb{R}^3\subset \mathbb{R}^5$
in space-time correspond to supersymmetric defects preserving
half of the supersymmetries (i.e. leading to ${\cal N}=2$ supersymmetry in $3d$).
We will be mainly considering non-compact Calabi-Yau threefolds which are toric.
A distinguished class of special Lagrangian cycles in these cases
\cite{mina1,mina2} have the topology
of $\mathbb{R}\times T^2$ for which a cycle of $T^2$ shrinks at each end.
 In the compact region of the toric 3-fold, where one cycle $w_0$ of $T^2$ shrinks
it ends on the web of the toric diagram.  With no loss of generality let us call this the $(1,0)$
cycle of $T^2$.   At infinity a cycle $w_\infty$
of the $T^2$ shrinks ending on the `spectators brane'.  Let us call this direction the $w_\infty =(p,q)$.
The topology of this Lagrangian submanifold is the lens space $L(q,p)$ which has fundamental
group ${\bf Z}_q$.  As discussed in \cite{CCV} there is an $ {\cal N}=2$ supersymmetric
$U(1)$ Chern-Simons gauge theory
living on the non-compact 3 dimensions of the wrapped M5 brane, with level $q$.  Furthermore this theory has a flavor $U(1)$ symmetry
associated with the monopole number (corresponding to shifting the angular scalar dual to the photon).  The $p$ corresponds to the Chern-Simons level for this flavor symmetry.
Furthermore the position of the brane on the web is determined by the FI-term $\xi_0$ for the $U(1)$
gauge symmetry.  Such a defect can be defined for any 5d conformal theory arising from
toric CY, where
the lines of the web pass through the same point.  As the web is resolved through
breathing modes of the web, one per cycle, the spectator $(q,p)$ line
where the brane is suspended can intersect a number of edges in the diagram
and the brane can end on any of the lines (see Fig \ref{lagg1}).  In order to make sure
the amplitudes is invariant under resolutions, and it is a defect
associated to a superconformal theory, we need to sum over all such possible endings.

In case the toric geometry engineers an $SU(N)$ gauge theory (corresponding
to $N$ parallel lines, the $(p,q)$ spectator line will intersect
the ladder of parallel lines at any of $N$ points, and we will need to sum
over all of them.  This would correspond to breaking $SU(N)$ to $SU(N-1)\times U(1)$
near the defect position.  Moreover as discussed in \cite{GW2} in the analogous
situation of surface defects in 4 dimensions, the surface defect generates a deficit angle
$0\leq \alpha\leq 2\pi $ in the $U(1)\subset U(1) \times SU(N-1)$, proportional to FI-term $\xi_0$. We have
$$\xi_0={\alpha \over 2\pi g_{YM}^2}$$
 corresponding to moving the end brane
along the line whose length is $1/g_{YM}^2$, as the brane traverses the line
the deficit angle varies from $0 $ to $2\pi$.  As we change the Coulomb branch parameters
the effective $\xi_{\pm}$ depends not only on the Coulomb parameter $a$ but also
on the slope $p/q$ (see Fig \ref{lagg1}).

In computing the index $I$ in the presence of defect we choose a number
of defect spectators with various slopes $(p_i,q_i)$ and some fixed positions (corresponding
to their FI-terms $\xi_i$).  We can also have more than one brane suspended from each.
In the gauge theory setup this will translate to more general patterns of breaking
the gauge symmetry near the defect.
We then integrate over the breathing modes of the loops (i.e. Wilson lines
of the 5d gauge theory), and the Wilson lines
associated to the gauge field on the brane, fixing the position of the
suspended lines at infinity and the external lines of the web, which collectively
play the role of mass parameters.

\section{Topological Strings and BPS states}

The $N=2$ topological strings propagating on a CY threefold $X$ have been intensely studied in recent years from both mathematical and physical viewpoints. They not only provide an exactly solvable sector of the full string theory but also provide very useful insight into the spacetime physics. In this section we will summarize the relation between topological strings on $X$ and BPS states which arise in the M-theory compactification on $X$.\\

\noindent Consider a Calabi-Yau threefold $X$ and let $\omega=\sum_{a=1}^{h^{1,1}(X)}t_{a}\omega_{a}$ be the K\"ahler class. The classes $\{\omega_{1},\omega_{2},\cdots,\omega_{h^{1,1}}\}$ span $H^{2}(X,\mathbb{Z})$ and we denote with $D_{a}$ the 4-cycle dual to $\omega_{a}$. The genus $g$ A-model topological string amplitude on the Calabi-Yau threefold $X$ are then given by \cite{BCOV}
\bea
F_{0}(\omega)&=&\frac{c_{abc}\,t_{a}t_{b}t_{c}}{6}+\sum_{\beta\in H_{2}(X,\mathbb{Z})}N^{0}_{\beta}e^{-\int_{\beta}\omega}\\\nn
F_{1}(\omega)&=&-\frac{1}{24}\sum_{a=1}^{h^{1,1}}t_{a}\int_{X}c_{2}(X)\wedge \omega_{a}\,+\sum_{\beta\in H_{2}(X,\mathbb{Z})}N^{1}_{\beta}e^{-\int_{\beta}\omega}\\\nn
F_{g\geq 2}(\omega)&=&(-1)^{g}\Big(\int_{{\cal M}_{g}}\lambda_{g-1}^{3}\Big)\frac{\chi(X)}{2}+\sum_{\beta\in H_{2}(X,\mathbb{Z})}N^{g}_{\beta}e^{-\int_{\beta}\omega}
\eea
where $c_{abc}=\int_{X}\omega_{a}\wedge \omega_{b}\wedge\omega_{c}$ are the triple intersection number $D_{a}\cdot D_{b}\cdot D_{c}$ of the divisors $D_{a}$ dual to $\omega_{a}$, $N^{g}_{\beta}$ are the genus $g$ Gromov-Witten invariants and the $\lambda_{g-1}$ is the $(g-1)th$ Chern class of the Hodge bundle over the moduli space of genus $g$ curves, ${\cal M}_{g}$, and
\bea
\int_{{\cal M}_{g}}\,\lambda^{3}_{g-1}=\frac{|B_{2g}||B_{2g-2}|}{(2g)(2g-2)(2g-2)!}\,.
\eea
In the above equation $B_{2g}$ are the Bernoulli numbers, $\sum_{n=0}^{\infty}B_{n}\frac{x^{n}}{n!}=\frac{t}{e^{t}-1}$.\\

\noindent The topological string partition function is given by
\bea
Z(\omega,g_{s})=\mbox{exp}\Big(\sum_{g=0}^{\infty}g_{s}^{2g-2}\,F_{g}(\omega)\Big)
\eea
where $g_{s}$ is the topological string coupling constant. In \cite{GV} topological strings on a CY threefold $X$ were studied from a spacetime point of view  and it was shown that the topological string partition function captures the degeneracy of BPS particles in the 5D theory coming from M-theory on $X$. We present a short summary of their argument linking the BPS states in 5D with topological strings. Consider M-theory compactification on CY threefold $X$ which gives a 5D theory. The massive BPS particles will form representation of the little group in 5D $SO(4)=SU(2)_{L}\times SU(2)_{R}$. These BPS particles in 5D arise from M2-branes wrapping a holomorphic curve in $X$ and have mass equal to the area of the curve. These BPS particles are electrically charged under the $h^{1,1}(X)$ abelian gauge fields $A^{(a)}$ coming from the 3-form $C$,
\bea
C=\sum_{a=1}^{h^{1,1}(X)}A^{(a)}\wedge\omega_{a}\,.
\eea
As has been mentioned before the 5D theory also has states which are magnetically charged under $A^{(a)}$. These magnetically charged states are not point particles but are strings coming from M5-branes wrapping the 4-cycles in $X$. The M2-brane wrapping a holomorphic curve in the class $\beta$ gives rise to a set of BPS particles in 5D with mass equal to $\int_{\beta}\omega$ and certain $SU(2)_{L}\times SU(2)_{R}$ spin content. Let use denote by $N^{j_{L},j_{R}}_{\beta}$ the number of particles with spin $(j_{L},j_{R})$ and charge $\beta$ (which determines the mass) and let
\bea
n_{\beta}^{j_{L}}=\sum_{j_{R}}(-1)^{2j_{R}}(2j_{R}+1)N^{j_{L},j_{R}}_{\beta}\,.
\eea
The integers $n^{j_{L}}_{\beta}$ are invariant under complex structure deformations of $X$ and are the BPS degeneracies captured by the topological strings. In terms of $n^{j_{L}}_{\beta}$ the topological string partition function can be written as $(q=e^{ig_{s}})$
\bea\nn
Z(\omega,g_{s})&=&Z_{0}(\omega,g_{s})\prod_{\beta\in H_{2}(X,\mathbb{Z})}\prod_{j_{L}}\prod_{k_{L}=-j_{L}}^{+j_{L}}\prod_{m=1}^{\infty}
\Big(1-q^{2k_{L}+m}\,e^{-\int_{\beta}\omega}\Big)^{m(-1)^{2j_{L}}n^{j_{L}}_{\beta}
}\\\nn
Z_{0}(\omega,g_{s})&=&\frac{\mbox{exp}\Big( \frac{c_{ijk}\,t_{i}t_{j}t_{k}}{6\,g_{s}^2}-\frac{1}{24}\sum_{a=1}^{h^{1,1}}t_{a}\int_{X}c_{2}(X)\wedge \omega_{a}\Big)}{\mbox{exp}\Big(-\frac{\zeta(3)}{g_{s}^2}+\sum_{g=2}^{\infty}g_{s}^{2g-2}(-1)^{g}\,\int_{{\cal M}_{g}}\lambda_{g-1}^{3}\Big)^{-\frac{\chi(X)}{2}}}
\eea
In $Z_{0}(\omega,g_{s})$ above the numerator is the classical contribution coming from worldsheet with genus zero and three punctures (the cubic term) the worldsheet with genus one and one puncture. The denominator is the contribution coming from constant maps and can also be written as
\bea
-\frac{\zeta(3)}{g_{s}^2}+\sum_{g=2}^{\infty}g_{s}^{2g-2}(-1)^{g}\,\int_{{\cal M}_{g}}\lambda_{g-1}^{3}=-\sum_{n=1}^{\infty}n\mbox{log}\Big(1-q^{n}\Big)=\mbox{log}\,M(q)
\eea
where $M(q)=\prod_{n=1}^{\infty}(1-q^{n})^{-n}$ is the generating function of the number of plane partitions known as MacMahon function. Thus the full topological string partition function is given by
\bea\nn
Z(\omega,g_{s})=e^{ \frac{c_{ijk}\,t_{i}t_{j}t_{k}}{6\,g_{s}^2}-\frac{1}{24}\sum_{a=1}^{h^{1,1}}t_{a}\int_{X}c_{2}(X)\wedge \omega_{a}}M(q)^{\frac{\chi(X)}{2}}\prod_{\beta,j_{L},k_{L},m}
\Big(1-q^{2k_{L}+m}\,e^{-\int_{\beta}\omega}\Big)^{m(-1)^{2j_{L}}n^{j_{L}}_{\beta}
}\,.
\eea
There also exist a refinement of the above topological string partition function. Notice that the GV invariants $n^{j_{L}}_{\beta}$ is an index over the Hilbert space of states coming from $\beta$ and the index structure is needed since complex structure deformations can change $N^{j_{L},j_{R}}_{\beta}$ but do not change $n^{j_{L}}_{\beta}$. This is the story for generic CY threefold. For local CY threefold (noncompact toric CY threefolds) the story is much more interesting. The local CY threefolds enjoy extra R symmetry and, therefore, $N^{j_{L},j_{R}}_{\beta}$ are also invariants. The refinement of topological string partition function captures these full BPS degeneracies \footnote{In the previous sections the coupling constants of the refined topological strings were denoted by $q_{1}$ and $q_{2}$. From now on we will denote them by $t$ and $q$ which are more familiar in the context of calculations involving the refined topological vertex.}:
\bea\label{rpf}
 &&Z(\omega,t,q)=e^{ -\frac{c_{ijk}\,t_{i}t_{j}t_{k}}{6\,\epsilon_{1}\epsilon_{2}}-\frac{1}{24}\sum_{a=1}^{h^{1,1}}t_{a}\int_{X}c_{2}(X)\wedge \omega_{a}}(M(t,q)M(q,t))^{\frac{\chi(X)}{4}}\times\\\nn
&&\prod_{\beta\in H_{2}(X,\mathbb{Z})}\prod_{j_{L},j_{R}}\prod_{k_{L}=-j_{L}}^{+j_{L}}\prod_{k_{R}=-j_{R}}^{+j_{R}}\prod_{m_{1},m_{2}=1}^{\infty}
\Big(1-t^{k_{L}+k_{R}+m_{1}-\frac{1}{2}}\,q^{k_{L}-k_{R}+m_{2}-\frac{1}{2}}\,Q^{\beta}\Big)^{M_{\beta}^{j_{L},j_{R}}}\\\nn
&&M_{\beta}^{j_{L},j_{R}}=(-1)^{2(j_{L}+j_{R})}N^{j_{L},j_{R}}_{\beta}\,,
\eea
where $M(t,q)$ is the refined MacMahon function,
\bea
M(t,q)=\prod_{i,j=1}^{\infty}\Big(1-q^{i}\,t^{j-1}\Big)^{-1}\,,
\eea
and $q=e^{i\epsilon_{1}},t=e^{-i\epsilon_{2}}$. The usual topological string partition function is recovered in the limit $\epsilon_{1}=-\epsilon_{2}=g_{s}$. Notice that we have kept the classical contribution and the constant map contribution.  Eq.(\ref{rpf}) can also be written as
\bea\nn
Z(\omega,t,q)&=&e^{ -\frac{c_{ijk}\,t_{i}t_{j}t_{k}}{6\,\epsilon_{1}\epsilon_{2}}-\frac{1}{24}\sum_{a=1}^{h^{1,1}}t_{a}\int_{X}c_{2}(X)\wedge \omega_{a}}(M(t,q)M(q,t))^{\frac{\chi(X)}{4}}\times \mbox{PE}\Big[F(\omega,t,q)\Big]\\\label{pe}
F(\omega,t,q)&=&\sum_{\beta\in H_{2}(X,\mathbb{Z})}\sum_{j_{L},j_{R}}\,e^{-\int_{\beta}\omega}\,\frac{(-1)^{2(j_{L}+j_{R})}N_{\beta}^{j_{L},j_{R}}
\mbox{Tr}_{j_{R}}(\frac{q}{t})^{j_{R,3}}\mbox{Tr}_{j_{L}}(q\,t)^{j_{L,3}}}{(q^{1/2}-q^{-1/2})(t^{1/2}-t^{-1/2})}\,
\eea
where $\mbox{PE}\Big[f(x_{1},x_{2},\cdots)\Big]$ is the Plethystic exponential of $f(x)$ defined as
\bea
\mbox{PE}\Big[f(x_{1},x_{2},\cdots)\Big]=\mbox{exp}\Big(\sum_{n=1}^{\infty}\frac{f(x_{1}^n,x_{2}^n,\cdots)}{n}\Big)\,.
\eea
In general for local CY threefold $X$ $\chi(X)$ is not well defined, however, if we only consider compact homologies in its definition we get it equal to twice the number of 4-cycles. This is the value we will use in writing the factors of MacMahon function in the refined partition functions.

\noindent Before we discuss how the refined topological string partition function can be calculated for local CY threefolds let us discuss an important property of the partition function which has been mentioned before and which will be of importance later. We would like to see how the refined partition function transforms under complex conjugation. In the later calculations of the index, as have been discussed earlier, the K\"ahler parameters will be taken to be pure imaginary and some of them will be integrated over. Keeping this in mind the complex conjugation acts as follows on the variables $(\omega,t,q)$,
\bea
(\omega,t,q)\mapsto (-\omega,t^{-1},q^{-1})\,.
\eea
Now it is easy to see from Eq.(\ref{pe}) that \footnote{As long as for each $\beta$ we have the full spin content corresponding to $(j_{L},j_{R})$. This is indeed the case for the class $\beta$ if the corresponding moduli space of D-brane ${\cal M}_{\beta}$ is compact. A counter example to this is the case of ${\cal O}(-2)\oplus {\cal O}(0)\mapsto \mathbb{P}^{1}$. In this case the moduli space of the $\mathbb{P}^{1}$ is $\mathbb{C}$ and the corresponding $F(T,t,q)=e^{-T}\,\frac{\sqrt{\frac{q}{t}}}{(q^{1/2}-q^{-1/2})(t^{1/2}-t^{-1/2})}$.}
\bea\label{sym}
F(-\omega,t^{-1},q^{-1})=F(-\omega,t,q)\,.
\eea
The MacMahon function, which is part of the closed topological string partition function, behaves in a non-trivial way under the complex conjugation,
\bea
M(t^{-1},q^{-1})&=&\prod_{i,j=1}^{\infty}\Big(1-q^{-i}\,t^{-j+1}\Big)^{-1}=\mbox{exp}\Big(\sum_{i,j=1}^{\infty}\sum_{n=1}^{\infty}\frac{q^{-ni}t^{-n(j-1)}}{n}\Big)\\\nn
&=&\mbox{exp}\Big(\sum_{n=1}^{\infty}\frac{q^{-n}}{n(1-q^{-n})(1-t^{-n})}\Big)=\mbox{exp}\Big(\sum_{n=1}^{\infty}\frac{t^{n}}{n(1-q^{n})(1-t^{n})}\Big)\\\label{sym2}
&=&\prod_{i,j=1}^{\infty}\Big(1-q^{i-1}\,t^{j}\Big)^{-1}=M(q,t)
\eea
Thus from Eq.(\ref{sym}) and Eq.(\ref{sym2}) it follows that
\bea\label{sym3}
\overline{Z(\omega,t,q)}=Z(-\omega,t^{-1},q^{-1})=Z(-\omega,t,q)
\eea
and therefore
\bea
\Big|Z(\omega,t,q)\Big|^2=\Big(M(t,q)M(q,t)\Big)^{\frac{\chi(X)}{2}}\,\mbox{PE}\Big[F(\omega,t,q)+F(-\omega,t,q)\Big]
\eea
where the classical piece cancelled because it was odd in $\omega$.\\

\noindent Now we will briefly discuss the open string case which will be of use when we consider the 5D index with a 3D defect. In the A-model topological string one can  consider worldsheet with boundaries as long as proper boundary conditions are enforced which preserve the A-model supersymmetry. The boundary conditions in this case require the boundary of the worldsheet to end on a Lagrangian submanifold of the target space. These Lagrangian submanifolds on which the worldsheet can have boundaries are the Lagrangian branes of the theory. For the local CY threefolds we are considering these Lagrangian branes are non-compact and have the topology of $S^{1}\times \mathbb{R}^{2}$. The partition function of the A-model in the presence of branes was studied in \cite{OV} from a spacetime viewpoint and it was shown that in this case, just as in the case of closed strings, the partition function captures certain BPS degeneracies. The spacetime picture arises if we consider Type IIA compactification and consider a D4-brane wrapped on the Lagrangian cycle.  In this D2-branes can wrap holomorphic curves in $X$ and end on the D4-brane. The open topological string partition function captures the degeneracies of BPS states arising from D2-branes ending on the D4-brane. If we denote the Lagrangian brane by ${\cal L}$ then the D4-brane wraps ${\cal L}\times {\mathbb R}^{2}$ where ${\mathbb R}^{2}$ is part of the spacetime ${\mathbb R}^{4}$. The theory on the $\mathbb{R}^2$ has a $U(1)_{s}$ rotation and a $U(1)_{r}$ R-symmetry. We combine these two $U(1)$'s and define $S_{L}=S+R$ and $S_{R}=S-R$. In addition to these quantum numbers the D2-brane couples to the gauge field on the D4-brane and we can introduce a holonomy factor $\mbox{Tr}_{R}U$ where $U$ is the holonomy of the gauge field on the D4-brane around the nontrivial $S^{1}$ of ${\cal L}$. If we denote by $N^{s_{L},s_{R}}_{R,\beta}$ the number of particles with charge $\beta$ and $U(1)_{L}\times U(1)_{R}$ quantum numbers $s_{L},s_{R}$ in the representation $R$, then the open topological string partition function is given by,
\bea
Z_{open}(\omega,t,q,U)&=&\mbox{PE}\Big[F_{open}(\omega,t,q,U)\Big]\\
F_{open}(\omega,q)&=&\sum_{R,\beta,s_{L}}e^{-\int_{\beta}\omega}\,(-1)^{2s_{L}}n^{s_{L}}_{R,\beta}
\frac{q^{s_{L}}}{(q^{\frac{1}{2}}-q^{-\frac{1}{2}})}\mbox{Tr}_{R}U
\eea
where
\bea
n^{s_{L}}_{R,\beta}=\sum_{s_{R}}N^{s_{L},s_{R}}_{\beta,R}(-1)^{2s_{L}+2s_{R}}
\eea
The $S_{L}+S_{R}$ is the fermion number and above index is invariant under complex structure deformations.

\noindent A refinement of the above partition function also exists and is given by \cite{GSV}
(see also \cite{AS})
\bea
Z_{open}(\omega,t,q)&=&\mbox{PE}\Big[F(\omega,t,q)\Big]\\\label{op}
F(\omega,t,q)&=&\sum_{\beta,R,s_{L},s_{R}}e^{-\int_{\beta}\omega}\,(-1)^{2s_{L}+2s_{R}}N^{s_{L},s_{R}}_{R,\beta}
\frac{q^{s_{L}}\,t^{s_{R}}}{(q^{\frac{1}{2}}-q^{-\frac{1}{2}})}\mbox{Tr}_{R}U
\eea

The action of complex conjugation on the open string partition function is different than in the case of the closed string partition function that we discussed above. The action of complex conjugation on the open string variables is given by
\bea
(\omega,t,q,U)\mapsto (-\omega, t^{-1},q^{-1},U^{-1})\,.
\eea
With this action the Eq.(\ref{op}) gives
\bea
F(-\omega,t^{-1},q^{-1})=-F(-\omega,t,q,U^{-1})\,,\\\nn
Z_{open}(-\omega,t^{-1},q^{-1},U^{-1})=\frac{1}{Z_{open}(-\omega,t,q,U^{-1})}\,.
\eea
Thus for the open string case
\bea
\Big|Z_{open}(\omega,t,q,U)\Big|^{2}=\frac{Z_{open}(\omega,t,q,U)}{Z_{open}(-\omega,t,q,U^{-1})}
\eea

\hskip-1cm \textcolor[rgb]{1.00,0.50,0.25}{\rule{6.45in}{0.02in}}\

\noindent {\bf An Example:} Consider the case of ${\cal O}(-1)\oplus {\cal O}(-1)\mapsto \mathbb{P}^{1}$ with a Lagrangian brane on the $\mathbb{P}^{1}$. In this case the open string partition function is given by
\bea\nn
Z_{open}(Q,t,q,z)= \prod_{m=1}^{\infty}\Big(1-q^{m-\frac{1}{2}}\,z\Big)^{-1}\Big(1-q^{m-\frac{1}{2}}Q\,z^{-1}\Big)^{-1}\,.
\eea
Using
\bea
\prod_{m=1}^{\infty}\Big(1-q^{-m+\frac{1}{2}}\,z\Big)^{-1}&=&\mbox{exp}\Big(\sum_{n=0}^{\infty}
\frac{z^{n}}{n}\frac{q^{-n/2}}{1-q^{-n}}\Big)\\\nn
&=&\mbox{exp}\Big(-\sum_{n=0}^{\infty}\frac{z^{n}}{n}\frac{q^{n/2}}{1-q^n}\Big)\\\nn
&=&\prod_{m=1}^{\infty}\Big(1-q^{m-\frac{1}{2}}\,z\Big)
\eea
it is easy to see that
\bea\nn
Z_{open}(Q^{-1},t^{-1},q^{-1},z^{-1})=\frac{1}{Z_{open}(Q^{-1},t,q,z^{-1})}\,.
\eea

\hskip-1cm \textcolor[rgb]{1.00,0.50,0.25}{\rule{6.45in}{0.02in}}\

\section{Five dimensional superconformal theories from toric Calabi-Yau threefolds}

In this section we briefly recall the class of 5D superconformal theories for which
our methods yield the corresponding index.  See \cite{vafa} and references
therein for more detail.

\begin{figure}[h]
  \centering
  \includegraphics[width=1.8in]{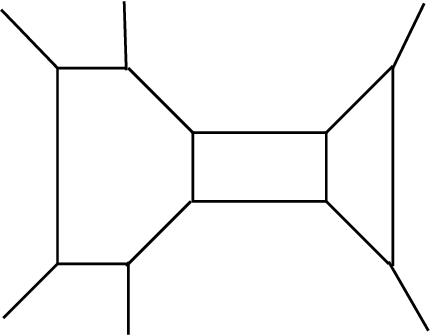}\\
  \caption{A generic $(p,q)$ 5-brane web.}\label{web}
\end{figure}

We consider M-theory on toric Calabi-Yau threefolds, or equivalently type IIB string theory with a web of $(p,q)$ 5-branes. Let $x^{0}, x^{1},\cdots, x^{9}$ be the  coordinates of the ten dimensional spacetime. The $(p,q)$ 5-branes fill the $\mathbb{R}^{1,4}$ part of the spacetime given by $x^{0},x^{1},\cdots x^{4}$ and extend as a web of piecewise straight lines in the plane given by $x^{5}$ and $x^{6}$. The generic $(p,q)$ 5-brane web can be viewed as a trivalent graph in $\mathbb{R}^2$ depicting each 5-branes as a line segment in $\mathbb{R}^2$
(filling the $\mathbb{R}^{1,4}$ space-time) where the slope of each $(p,q)$ line is given by the $q/p$. The generic graph is trivalent with $\sum_i (p_i,q_i)=0$ on each vertex. An example of such a web is shown in \figref{web}.

\begin{figure}[h]
  \centering
  \includegraphics[width=2in]{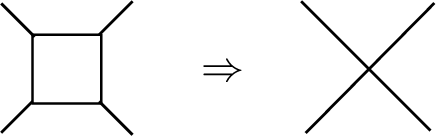}\\
  \caption{The singular limit of the web gives a superconformal theory. In this case the theory has $SU(2)$ global symmetry at the superconformal point. In the M-theory compactification this corresponds to a 4-cycle ($\mathbb{P}^{1}\times \mathbb{P}^{1}$) shrinking to a point.}\label{superconformal}
\end{figure}

In the limit where the web becomes singular, consisting of lines all passing through the same point, we get a superconformal theory in 5D. An example is shown in \figref{superconformal} where the singular web gives a superconformal theory with $SU(2)$ global symmetry.

The resolutions of the web, fixing the external line, correspond to going
to the Coulomb branch of the 5D gauge theory.  Some of these theories correspond to gauge theories upon resolutions  \cite{peter}.  However most of them do not have a direct gauge theory interpretation. Our method for computing the index applies equally well to all of them.

Moving the external lines, correspond to changing the mass parameters of the theory. The data of the conformal theory is thus captured by a collection of external lines characterized by ${\bf w}_i=(p_i,q_i)$
5-branes, with the condition that
$$\sum_{i}{\bf w}_i =0$$
Moreover, for each $\vec{w}_i$ one can introduce a mass parameter $m_i$ corresponding
to moving the external lines parallel to itself.  They add up to zero and there is in addition
a two parameter redundancy due to shifting the origin of the $\mathbb{R}^2$, so the number of
mass parameters is 3 less than the number of external lines.
It was proposed in \cite{vafa} that
this data can be identified with the states of a 4D string on $T^* T^2$.  Moreover, the
scattering amplitudes of the resulting string states are identified with the
superconformal index $I_5$ of the resulting theory in 5D:
$$\langle \prod_i \Phi_{{\bf w}_i}(m_i)\rangle =I_5 \ \delta (\sum {m_i})\  \delta (\sum {\bf w}_i)$$
In addition we can select a number of
spectator branes from which the Lagrangian branes can be suspended, giving
rise to defects of the 5d theory.  The slope of the spectator branes determine
the type of defect we introduce.  Its position is a mass parameter associated to the FI-term
on the defect.  These correspond to degrees of freedom of the unwound string in the
proposal of \cite{vafa}.

\subsection{Loop variables and K\"ahler parameters}
In calculating the index we need to integrate over the loop variables associated with the 4-cycles in the geometry. Each loop variable correspond to a $U(1)$ coming from the 4-cycle as discussed in section 4. Since the partition function depends on the K\"ahler parameters we need to determine how the K\"ahler parameters depend on the loop variables. This relation can be easily determined either from the web diagram or from the geometry.

\begin{figure}[h]
  \centering
  \includegraphics[width=2in]{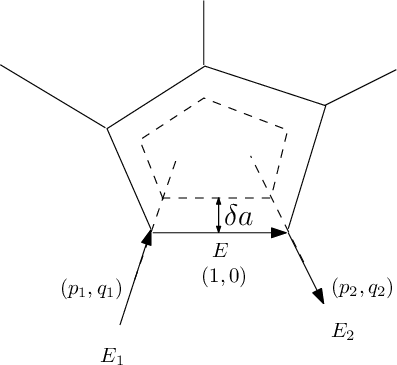}\\
  \caption{}\label{loopvar}
\end{figure}

Let us first show how we can determine the dependence of the K\"ahler parameters on the loop variables using the web diagram. Consider an edge $E$ which is one of the edges forming the loop (4-cycle) in the web diagram. Let $E_{1}$ and $E_{2}$ be the two edges connected with $E$ but not part of the loop as shown in \figref{loopvar} where we have used $SL(2,\mathbb{Z})$ transformation to convert the edge $E$ to a horizontal line.

\noindent From \figref{loopvar} it is clear that as the 4-cycle size changes the size of the edge $E$ also changes with it. The relation between the deformation of the 4-cycle given by the change in the loop variable $\delta\,a$ and the change in the size of the edge $E$ $\delta t_{E}$ depends on the slope of the connected edges $E_{1}$ and $E_{2}$ and is given by
\bea
\delta t_{E}=\delta a\Big|\frac{p_{1}}{q_{1}}-\frac{p_{2}}{q_{2}}\Big|\,.
\eea
If we define $Q_{e}=e^{it_{E}}$, the loop variable $U=e^{ia}$ and let $w_{1}, w_{2}$ and $w_{e}$ be the winding vectors associated with $E_{1}, E_{2}$ and $E$ then the $SL(2,\mathbb{Z})$ invariant version of the relation between the edge variable $Q_{e}$ and the loop variable $U$ is given by
\bea
Q_{e}=Q_{0}\,U^{n}\,,\,\,\,\,n=\Big|\frac{w_{1}\wedge w_{2}}{(w_{1}\wedge w_{e})\,(w_{2}\wedge w_{e})}\Big|
\eea
where $Q_{0}$ is the value of $Q_{e}$ for $a=0$ and is determined by the position of the external legs.

If the geometry has many 4-cycles then it may becomes difficult to determine the dependence of the K\"ahler parameters on the loop variables using the web diagram although the basic idea still is same. A more geometric way of obtaining the relation follows from the fact that holomorphic curves in the geometry give rise to BPS particles in the 5D theory which are electrically charged under the $U(1)^g$ gauge group (assuming there are $g$ 4-cycles in the geometry). The scaling relation between the K\"ahler parameter of a curve $C$ and the loop variables is just given by the electric charge of the corresponding state:
\bea\label{scale}
Q_{C}=Q_{C,0}\,e^{i(d_{1}a_{1}+d_{2}a_{2}+\cdot d_{g}a_{g})}\,,
\eea
where $\{a_{1},a_{2},\cdots,a_{g}\}$ are the loop variables corresponding to the $g$ 4-cycles and $d_{i}$ is electric charge of the state coming from $C$ under the $U(1)_{i}$ (the $U(1)$ coming from the i-th 4-cycle). The electric charge of the curve $C$ is a purely geometric quantity given by the intersection of the curve $C$ with the 4-cycle. If we denote the 4-cycles in the geometry by $D_{1},D_{2},\cdots,D_{g}$ then
\bea\label{charge}
d_{i}(C)&=&D_{i}\cdot C\,\\
&=&-K_{D_{i}}\cdot C
\eea
where $-K_{D_{i}}$ is the anticanonical class of the divisor $D_{i}$. We will use Eq.(\ref{scale}) and Eq.(\ref{charge}) to determine the relation between the K\"ahler parameters and the loop variables when calculating the index in section 6.

\section{Computation of the 5d index through topological string}
In this section we will calculate the index for certain 5D theories using the refined topological string partition function. The refined partition function will be calculated using the refined topological vertex. We will give a short introduction to the refined vertex formalism

\subsection{Refined vertex formalism}
The topological vertex, which was derived using large N transition from Chern-Simons theory, can be used to calculate the topological string partition function for a toric CY threefold \cite{tv}. A refinement of the topological vertex was found in \cite{IKV} and allows the calculation of refined topological string partition function for a large class of toric CY threefolds \footnote{See \cite{awata1} for an earlier attempt at refining the topological vertex by replacing Schur polynomials with Macdonald polynomials.}.  The refined topological vertex is given by
\bea\nn
C_{\lambda\,\mu\,\nu}(t,q)&=&f_{\mu^t}(q,t)\,q^{\frac{||\nu||^2}{2}}\,\widetilde{Z}_{\nu}(t,q)
\sum_{\eta}\Big(\frac{q}{t}\Big)^{\frac{|\eta|+|\lambda|-|\mu|}{2}}\,s_{\lambda^{t}/\eta}(t^{-\rho}\,q^{-\nu})\,s_{\mu/\eta}(t^{-\nu^t}\,q^{-\rho})\\
\eea
where $s_{\lambda/\eta}(\mathbf{x})$ is the skew-Schur function and the following table summarizes other quantities:

\hskip-1cm \textcolor[rgb]{1.00,0.50,0.25}{\rule{6.45in}{0.02in}}\
\begin{doublespace}
$\lambda=\{\lambda_{1}\geq\lambda_{2}\geq \cdots \geq \lambda_{\ell(\lambda)}>0\}\,,\,\,\,\lambda^t=\{\lambda^t_{1}\geq\lambda^t_{2}\geq \cdots |\lambda^t_{i}=\#\{a|\lambda_{a}\geq i\}\}$\\
$|\lambda|=\sum_{a=1}^{\ell(\lambda)}\lambda_{a}\,,\,\,\,||\lambda||^2=\sum_{a=1}^{\ell(\lambda)}(\lambda_{a})^2$\\
$f_{\lambda}(t,q)=(-1)^{|\lambda|}\,t^{\frac{||\lambda^t||^2}{2}}\,q^{-\frac{||\lambda||^2}{2}}\,,\,\,\,\,\,\,
\widetilde{Z}_{\lambda}(t,q)=\prod_{i=1}^{\ell(\lambda)}\prod_{j=1}^{\lambda_{i}}\Big(1-q^{\lambda_{i}-j}\,t^{\lambda^t_{j}-i+1}\Big)^{-1}$\\
$\rho=\{-\frac{1}{2},-\frac{3}{2},-\frac{5}{2},\cdots\}\,,\,\,t^{-\rho}q^{-\lambda}=\{t^{\frac{1}{2}}\,q^{-\lambda_{1}},
t^{\frac{3}{2}}\,q^{-\lambda_{2}},t^{\frac{5}{2}}\,q^{-\lambda_{3}},\cdots\}$\\
$s_{\lambda/\mu}({\bf x})=\sum_{\eta}N^{\lambda}_{\mu\,\eta}s_{\eta}({\bf x})\,,\,\,\,N^{\lambda}_{\mu\,\eta}=\mbox{Littlewood-Richardson coefficients}$

\end{doublespace}
\hskip-1cm \textcolor[rgb]{1.00,0.50,0.25}{\rule{6.45in}{0.02in}}\

Given any web diagram corresponding to a toric Calabi-Yau threefold we give orientation to edge and associate to each internal edge $e_{\alpha}$ a partition $\lambda^{(\alpha)}$. To each external edge we associate the trivial partition i.e., the empty set. Since in the web diagram three edges meet at each vertex we have a set of three partitions for each vertex. If an edge is oriented such that it is going out from the vertex the corresponding partition is changed to its transpose. We use these three partitions, say $\lambda,\mu ,\nu$ associated to the incoming edges of the vertex, to associate with the vertex the refined topological vertex $C_{\lambda\,\mu\,\nu}(t,q)$. The ordering of the three partitions in writing the refined vertex is taken to be anticlockwise as we go around the vertex and this should be the same for all vertices in the web diagram.  To each edge $e_{\alpha}$ of the web diagram we had associated a partition $\lambda^{(\alpha)}$ and we now associate a factor of $e^{-|\lambda^{(\alpha)}|(t_{\alpha}+i\pi)}\,(f_{\lambda^{(\alpha)}}(t,q))^{p_{\alpha}}$ where $t_{\alpha}$ is the length of this edge $e_{\alpha}$ and $p_{\alpha}$ is an integer which is determined by the local geometry of the $\mathbb{P}^{1}$ associated to the edge $e_{\alpha}$ in the CY threefold. In the neighborhood of a $\mathbb{P}^{1}$ in a CY threefold the geometry looks like ${\cal O}(m_{1})\oplus {\cal O})(m_{2})$ with $m_{1}+m_{2}=-2$, the integer $p=(m_{2}-m_{1})/2$. Another important constraint that needs to be considered in the case of refined topological vertex, but not for the usual topological vertex, is that at each vertex we need to assign one edge as the preferred edge and all preferred edges in the web diagram should be parallel to each other. This constraint comes from the construction of the refined topological vertex in terms of plane partitions and restricts the class of toric CY threefolds to which refined vertex can be applied to those geometries which are fibrations over a $\mathbb{P}^{1}$ or a chain of $\mathbb{P}^{1}$'s. In writing the refined vertex for an vertex, of the web diagram, the partition associated with the preferred edge is always the last partition in the refined vertex and the two refined vertex factors which appear for two vertices connected by a preferred edge should have $(t,q)$ parameters switched between them. With these constraints in place the refined topological string partition function is given by taking the product over all vertices of the corresponding refined vertex factors and taking a product over all edges of the corresponding edge factors and summing over all partitions:
\bea\nn
Z_{refined}(t_{\alpha},t,q):=\sum_{\mbox{all\,\,partitions}}\prod_{\alpha}\Big(e^{-|\lambda^{(\alpha)}|(t_{\alpha}+i\pi)}\,(f_{\lambda^{(\alpha)}}(t,q))^{p_{\alpha}}\Big)
\prod_{vertices}C_{\lambda^{(\alpha)}\,\lambda^{(\beta)}\,\lambda^{(\gamma)}}
\eea

\hskip-1cm \textcolor[rgb]{1.00,0.50,0.25}{\rule{6.45in}{0.02in}}\

{\bf An Example:} Consider local $\mathbb{F}_{m}$, canonical bundle on Hirzebruch surface $\mathbb{F}_{m}$. The web diagram of this geometry is shown in figure below. We take the two horizontal lines to be the preferred edges.
\begin{figure}[h]
  \centering
  \includegraphics[width=3in]{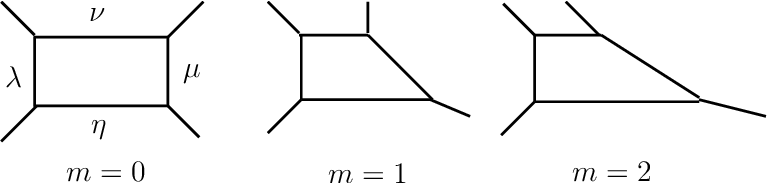}\\
\end{figure}
\bea
\mbox{Edge factor}:\,\,\,\,\,\,&&(-1)^{|\lambda|+|\mu|+|\nu|+|\eta|}\,
e^{-t_{f}(|\lambda|+|\mu|)+t_{b}|\nu|+(t_{b}+mt_{f})|\eta|}\,\times\\\nn
&&\,f_{\nu}(t,q)^{-m+1}\,f_{\eta}(q,t)^{m+1}\,f_{\lambda}(t,q)\,f_{\mu}(q,t)\\\nn
\mbox{Vertex factors}:\,\,\,\,\,\,&&C_{\lambda^t\emptyset\nu}(t,q)\,C_{\emptyset\,\lambda\,\eta^t}(t,q)
C_{\mu^t\emptyset\eta}(q,t)\,C_{\emptyset\,\mu\,\nu^{t}}(q,t)
\eea
 The refined partition function is then given by
 \bea
 Z_{local\,\,\mathbb{F}_{m}}&=&\sum_{\lambda\,\mu\,\nu^{(1)}\,\nu^{(2)}}\,\mbox{Edge factor}\times \mbox{Vertex factor}
 \eea
 After some simplification and using the identity $\sum_{\lambda}s_{\lambda}({\bf x})\,s_{\lambda}({\bf y})=\prod_{i,j}(1-x_{i}y_{j})^{-1}$ we get $(Q_{b}=e^{-t_{b}}\,,\,Q_{f}=e^{-t_{f}})$
 \bea\label{examplefm}
 Z_{local\,\,\mathbb{F}_{m}}&=&\sum_{\nu\,\eta}((-1)^{m}\,Q_{b})^{(|\nu+|\eta|)}Q_{f}^{m|\eta|}
(f_{\nu^t}(q,t))^{m}\,(f_{\eta}(q,t))^{m}\,q^{||\eta^t||^2}\,t^{||\nu^t||^2}\\\nn
&& \widetilde{Z}_{\nu}(t,q)\widetilde{Z}_{\eta^t}(t,q)\widetilde{Z}_{\eta}(q,t)\widetilde{Z}_{\nu^t}(q,t)\\\nn
&&
 \prod_{i,j=1}^{\infty}\Big[(1-Q_{f}\,t^{i-\eta_{j}}q^{j-1-\nu_{i}})(1-Q_{f}\,q^{i-\nu_{j}}t^{j-1-\eta_{i}})\Big]^{-1}
 \eea

 \hskip-1cm \textcolor[rgb]{1.00,0.50,0.25}{\rule{6.45in}{0.02in}}\

\subsection{Example 1: Local $\mathbb{P}^{1}\times \mathbb{P}^{1}$}
Let us begin with a very interesting example of local $\mathbb{P}^{1}\times \mathbb{P}^{1}$. This CY threefold gives rise to $N_{f}=0$ $SU(2)$ gauge theory and we will be able to compare the answer we get from topological strings with the gauge theoretic calculation of \cite{KKL}.\\

\noindent The web diagram corresponding to this CY threefold (which is dual to the Newton polygon encoding the toric data of this CY threefold) is shown below in Fig(\ref{figlocalp1p1}).

\begin{figure}[h]
  \centering
  \includegraphics[width=4in]{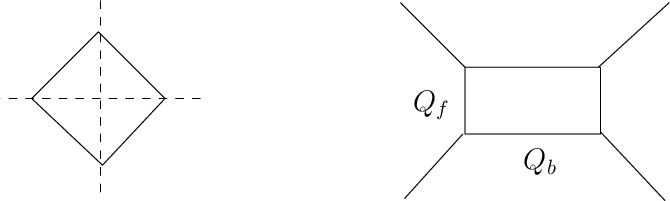}\\
 \caption{The Newton polygon (a) and web diagram (b) of local $\mathbb{P}^{1}\times \mathbb{P}^{1}$. The Newton polygon has a unique triangulation therefore this geometry has only one phase. }\label{figlocalp1p1}
\end{figure}
\noindent In the above figure $Q_{b}$ and $Q_{f}$ are related to the K\"ahler parameters $t_{b}$ and $t_{f}$ corresponding to the base $\mathbb{P}^{1}$, which we will denote by $B$, and the fiber $\mathbb{P}^{1}$, which we will denote by $F$, respectively as
\bea\nn
Q_{b}=e^{-t_{b}}\,,\,\,\,\,\,Q_{f}=e^{-t_{f}}\,.
\eea

\noindent The refined partition function of this geometry was calculated above and is given by taking $m=0$ in Eq.(\ref{examplefm}),
\bea\label{factors}
Z_{\tiny \mbox{\tiny local}\, \mathbb{P}^{1}\times \mathbb{P}^{1}}(Q_{b},Q_{f},t,q)&=&\Big(M(t,q)M(q,t)\Big)^{\frac{1}{2}}\,Z(Q_{b},Q_{f},t,q)\\\nn
Z(Q_{b},Q_{f},t,q)&:=&\sum_{\nu_{1}\,\nu_{2}}Q_{b}^{|\nu_{1}|+|\nu_{2}|}\,\,q^{||\nu_{2}^{t}||^2}\,t^{||\nu_{1}^{t}||^2}\widetilde{Z}_{\nu_{1}}(t,q)\widetilde{Z}_{\nu_{2}}(q,t)
\widetilde{Z}_{\nu_{1}^t}(q,t)\widetilde{Z}_{\nu_{2}^t}(t,q)\times \\\nn
&&\prod_{i,j=1}^{\infty}
\Big[\Big(1-Q_{f}t^{i-1-\nu_{2,j}}q^{j-\nu_{1,i}}\Big)\Big(1-Q_{f}q^{i-1-\nu_{1,j}}t^{j-\nu_{2,i}}\Big)\Big]^{-1}
\eea
The refined topological vertex calculation gives the last factor in Eq.(\ref{factors}). The first factor involving the refined MacMahon function $M(t,q)$ has been added in accordance with Eq.(\ref{rpf}) while taking $\chi(X)=2$, as discussed in section 5, since there is only one 4-cycles. We have ignored the classical contribution in writing the refined partition function since it cancels when we take the absolute value square of the refined partition function as discussed in section 5.\\

\begin{figure}[h]
  \centering
  \includegraphics[width=4in]{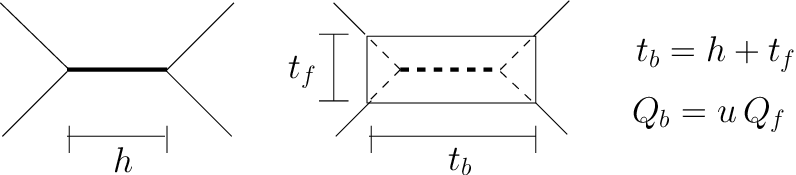}\\
  \caption{The parameter $h$ is determined by the position of the external legs and is fixed. In the 5D gauge theory $h$ is proportional to the inverse of the tree level gauge coupling and has dimensions of mass.}\label{loopp1p1}
\end{figure}

\noindent The the index for this geometry is given by
\bea\nn
I&=&\int da\,\Big|Z_{\tiny \mbox{\tiny local}\, \mathbb{P}^{1}\times \mathbb{P}^{1}}(Q_{b},Q_{f},t,q)\Big|^2
\eea
where $a$ is the loop variable (breathing mode) for the 4-cycle in the geometry. In section 6 we discussed the general relation between the K\"ahler parameters and the loop variable. In this case we see that the $Q_{f}$ is related to loop variable as
\bea
Q_{f}=e^{2i\,a}\,.
\eea
$Q_{b}$ also depends on the loop variable and, therefore, on $Q_{f}$. This dependence can be easily determined using the web diagram. Consider the web diagram shown in \figref{figlocalp1p1}. If the external legs are fixed then the two parameters $Q_{b}$ and $Q_{f}$ are not independent anymore, instead the choice of the external legs determines a parameter $u=e^{-h}$ such that $\frac{Q_{b}}{Q_{f}}=u$ as shown in the \figref{loopp1p1}.

\noindent Thus the index is given by,
\bea\nn
I_{\tiny \mbox{\tiny local}\, \mathbb{P}^{1}\times \mathbb{P}^{1}}&=&\int da\,\Big|Z_{\tiny \mbox{\tiny local}\, \mathbb{P}^{1}\times \mathbb{P}^{1}}(u\,e^{2ia},e^{2ia},t,q)\Big|^2\,.\\\nn
\eea
We can now use Eq.(\ref{factors}) to determine the above index to obtain \footnote{The index can also be written as a infinite product
\bea
I=\prod_{a,b,c}\Big(1-x^{a}y^{b}u^{c})^{C(a,b,c)}\,,
\eea
where $C(a,b,c)\in \mathbb{Z}$. It would be interesting to see if $C(a,b,c)$ have a direct physical meaning.}:

\bea\nn\label{indexp1p1}
I_{\tiny \mbox{\tiny local}\, \mathbb{P}^{1}\times \mathbb{P}^{1}}&:=&1+\chi_{3}(u)\,x^2 + \chi_{2}(y)\Big(1 + \chi_{3}(u)\Big)x^3 + \Big(\chi_{3}(y)\Big[1 + \chi_{3}(u)\Big] + 1 +
    \chi_{5}(u)\Big) x^4 \\\nn
    &&+\Big(\chi_{4}(y)\Big[1 + \chi_{3}(u)\Big] +
    \chi_{2}(y) \Big[1 + \chi_{3}(u) + \chi_{5}(u)\Big]\Big) x^5 +\\\nn
    &&\Big(\chi_{5}(y) \Big[1 + \chi_{3}(u)\Big] +
    \chi_{3}(y) \Big[1 + \chi_{3}(u) + \chi_{5}(u) + \chi_{3}(u) \chi_{3}(u)\Big] +\\\nn
    &&\chi_{3}(u) + \chi_{7}(u) - 1\Big) x^6 +\Big(\chi_{6}(y) \Big[1 + \chi_{3}(u)\Big] +\\\nn
&& \chi_{4}(y)\Big[2 + 4 \chi_{3}(u) + 2 \chi_{5}(u)\Big] +
    \chi_{2}(y) \Big[1 + 3 \chi_{3}(u) + 2 \chi_{5}(u) + \chi_{7}(u)\Big]\Big) x^7 +\\\nn
    &&
 \Big(\chi_{7}(y) \Big[1 + \chi_{3}(u)\Big] +
    \chi_{5}(y) \Big[4 + 5 \chi_{3}(u) + 3 \chi_{5}(u)\Big] +\\\nn
&&\chi_{3}(y) \Big[2 + 7 \chi_{3}(u) + 3 \chi_{5}(u) + 2 \chi_{7}(u)\Big] +3 +
    2 \chi_{3}(u) + 2\chi_{5}(u) +
    \chi_{9}(u)\Big) x^8 + \\\nn
    &&\Big(\chi_{8}(y)\Big[1 + \chi_{3}(u)\Big] +
    \chi_{6}(y)\Big[3 \chi_{5}(u) + 7\chi_{3}(u) + 4\Big] +\\\nn
&& \chi_{4}(y)\Big[3 \chi_{7}(u) + 6 \chi_{5}(u) + 10 \chi_{3}(u) + 6\Big] +\\\nn
    &&
   \chi_{2}(y)\Big[\chi_{9}(u) + 2 \chi_{7}(u) + 4 \chi_{5}(u) +
       7\chi_{3}(u) + 4\Big]\Big)x^9 + \Big(\chi_{9}(y)\Big[1 + \chi_{3}(u)\Big] +\\\nn
       &&
    \chi_{7}(y)\Big[4 \chi_{5}(u) + 8 \chi_{3}(u)+ 6\Big] +
    \chi_{5}(y)\Big[4 \chi_{7}(u) + 9\chi_{5}(u) + 16\chi_{3}(u) + 7\Big] +\\\nn
    &&
    \chi_{3}(y)\Big[2 \chi_{9}(u) + 4\chi_{7}(u) + 10\chi_{5}(u) +
       11\chi_{3}(u) + 10\Big] + \\
       &&\chi_{11}(u) + 3\chi_{7}(u) +3\chi_{5}(u) + 7\chi_{3}(u) + 1\Big)x^{10}+\cdots\,,
\eea
where $x=\sqrt{\frac{q}{t}}$ and $y=\sqrt{q\,t}$.\\

\noindent Eq.(\ref{indexp1p1}) agrees will the result of \cite{KKL} for the case of $SU(2)$ gauge theory with $N_{f}=0$. \\

\noindent In order to understand the relation between the gauge theoretic calculation \cite{KKL} and topological string result that we just derived we will look carefully at the various factors which arise in the calculation of the index.\\

In \cite{KKL} the index for $SU(2)$ gauge theory with $N_{f}=0$ was calculated using equivariant localization and was
given by
\bea\label{kim}
I=\int da\,\underbrace{2\mbox{sin}^{2}(a)\mbox{PE}\Big[f_{vec}(a,x,y)\Big]}_{perturbative\,contribution}\,
\underbrace{\Big|Z_{Nekrasov}(a,q,x,y)\Big|^2}_{instanton\,contribution}\,,
\eea
where $a$ is the parameter on the Coulomb branch, $q$ is the instanton counting parameter and $x$ and $y$ are related to the equivariant parameters $q$ and $t$ for the $U(1)\times U(1)$ action of $\mathbb{C}^{2}$,
\bea
(z_{1},z_{2})\in \mathbb{C}^{2}\mapsto (q\,z_{1},t^{-1}z_{2})\,,\\\nn
x=\sqrt{\frac{q}{t}}\,,\,\,\,y=\sqrt{q\,t}.\
\eea
The perturbative contribution after subtracting the Haar measure is given by $f_{vec}(a,x,y)$,
\bea\label{fvec}
f_{vec}(a,x,y)=-\frac{x(y+\frac{1}{y})}{(1-x\,y)(1-\frac{x}{y})}\,\Big(e^{2ia}+1+e^{-2ia}\Big)
\eea

\noindent Now we can identify different pieces of the integrand in Eq.(\ref{kim}) with different contribution to the
topological string partition function. The topological string partition function of local $\mathbb{P}^{1}\times \mathbb{P}^{1}$ can be written as
\bea\label{factors2}
Z_{\tiny \mbox{\tiny local}\, \mathbb{P}^{1}\times \mathbb{P}^{1}}(Q_{b},Q_{f},t,q)&=&\Big(M(t,q)M(q,t)\Big)^{\frac{1}{2}}\,Z_{0}(Q_{f},t,q)\,Z'(Q_{b},Q_{f},t,q)\\\nn
Z_{0}(Q_{f},t,q)&=&\prod_{i,j=1}^{\infty}\Big[\Big(1-Q_{f}\,q^{i}t^{j-1}\Big)\Big(1-Q_{f}q^{i-1}t^{j}\Big)\Big]^{-1}\\\nn
Z'(Q_{b},Q_{f},t,q)&=&\sum_{\nu_{1}\,\nu_{2}}Q_{b}^{|\nu_{1}|+|\nu_{2}|}\,\,q^{||\nu_{2}^{t}||^2}\,t^{||\nu_{1}^{t}||^2}\widetilde{Z}_{\nu_{1}}(t,q)\widetilde{Z}_{\nu_{2}}(q,t)
\widetilde{Z}_{\nu_{1}^t}(q,t)\widetilde{Z}_{\nu_{2}^t}(t,q)\times \\\label{nekins}
&&\prod_{i,j=1}^{\infty}\frac{(1-Q_{f}t^{i-1}q^{j})(1-Q_{f}q^{i-1}t^{j})}
{(1-Q_{f}t^{i-1-\nu_{2,j}}q^{j-\nu_{1,i}})(1-Q_{f}q^{i-1-\nu_{1,j}}t^{j-\nu_{2,i}})}
\eea

\noindent In Eq.(\ref{factors2}) $Z_{0}(Q_{f},t,q)$ is the contribution to the partition function coming from branes wrapping the fiber curve $F$ only and $Z'(Q_{b},Q_{f},t,q)$ is the contribution to the partition function coming from branes wrapping the base curve $B$ at least once and wrapping the fiber curve arbitrary number of times. The contribution $Z'(Q_{b},Q_{f},t,q)$ is such that
\bea
\lim_{Q_{b}\mapsto 0}\,\,Z'(Q_{b},Q_{f},t,q)=1
\eea
Thus in the limit $Q_{b}\mapsto 0$  the only contribution to the partition function comes from branes wrapping the fiber curve $F$ and the D0-branes (the constant map contribution).
$Z_{0}(Q_{f},t,q)$ gives the perturbative part of the 4D gauge theory partition function in the limit
\bea
Q_{f}=e^{2ia\beta}\,,\,\,q=e^{\beta\epsilon_{1}}\,\,t=e^{-i\beta\epsilon_{2}}\,,\,\,\beta\mapsto 0\,.
\eea
The index is expressed in terms of the variables $x$ and $y$ which couple to the $SU(2)_{R}$ and $SU(2)_{L}$ spins. To see the relation between the integrand of the index and the topological string partition function lets express partition function $\Big|M(t,q)Z_{0}(Q_{f},t,q)\Big|^2$ in terms of the variables $x=\sqrt{\frac{q}{t}}$ and $y=\sqrt{q\,t}$:
\bea
M(t,q)&=&\prod_{i,j=1}^{\infty}\Big(1-q^{i}t^{j-1}\Big)^{-1}=\prod_{i,j=1}^{\infty}\Big(1-x^{i+j}y^{i-j}\Big)\\\nn
\overline{M(t,q)}&=&M(t^{-1},q^{-1})=M(q,t)=\prod_{(i,j)\neq (1,1)}^{\infty}\Big(1-x^{i+j-2}y^{i-j}\Big)\\\nn
|M(t,q)|^{2}&=&M(t,q)M(q,t)=\prod_{i,j=1}^{\infty}\Big(1-x^{i+j}y^{i-j}\Big)\prod_{(i,j)\neq (1,1)}^{\infty}\Big(1-x^{i+j-2}y^{i-j}\Big)\\\label{g1}
&=&\prod_{i,j=1}^{\infty}\Big(1-x^{i+j-1}y^{i-j+1}\Big)\Big(1-x^{i+j-1}y^{i-j-1}\Big)\\\nn
&=&\Big|\prod_{i,j=1}^{\infty}\Big(1-x^{i+j-1}y^{i-j+1}\Big)\Big|^{2}
\eea
Similarly
\bea\label{g2}
Z_{0}(Q_{f},t,q)&=&\Big[\prod_{i,j=1}^{\infty}\Big(1-Q_{f}q^{i}t^{j-1}\Big)\Big(1-Q_{f}t^{i}q^{j-1}\Big)\Big]^{-1}\\\nn
&=&\prod_{i,j=1}^{\infty}\Big(1-Q_{f}x^{i+j}y^{i-j}\Big)\Big(1-Q_{f}x^{i+j-2}y^{i-j}\Big)\\\nn
&=&(1-Q_{f})\prod_{i,j=1}^{\infty}\Big(1-Q_{f}x^{i+j}y^{i-j}\Big)\prod_{(i,j)\neq (1,1)}\Big(1-Q_{f}x^{i+j-2}y^{i-j}\Big)\\\nn
&=&(1-Q_{f})\prod_{i,j=1}^{\infty}\Big(1-Q_{f}x^{i+j-1}y^{i-j+1}\Big)\prod_{i,j=1}^{\infty}\Big(1-Q_{f}x^{i+j-1}y^{i-j-1}\Big)
\eea
\bea\label{g3}
\overline{Z_{0}(Q_{f},t,q)}&=&\prod_{i,j=1}^{\infty}\Big(1-Q_{f}^{-1}x^{-i-j}y^{-i+j}\Big)\Big(1-Q_{f}^{-1}x^{-i-j+2}y^{-i+j}\Big)\\\nn
&=&\prod_{i,j=1}^{\infty}\Big(1-Q_{f}^{-1}x^{i+j-2}y^{i-j}\Big)\Big(1-Q_{f}^{-1}x^{i+j}y^{i-j}\Big)\\\nn
&=&(1-Q_{f}^{-1})\prod_{i,j=1}^{\infty}\Big(1-Q_{f}^{-1}x^{i+j-2}y^{i-j}\Big)\prod_{(i,j)\neq (1,1)}\Big(1-Q_{f}^{-1}x^{i+j}y^{i-j}\Big)\\\nn
&=&(1-Q_{f}^{-1})\prod_{i,j=1}^{\infty}\Big(1-Q_{f}^{-1}x^{i+j-1}y^{i-j+1}\Big)\prod_{i,j=1}^{\infty}\Big(1-Q_{f}^{-1}x^{i+j-1}y^{i-j-1}\Big)
\eea

\noindent Using Eq.(\ref{fvec}) it is easy to see that
\bea\label{indexpert}
&&2\mbox{sin}^{2}(a)PE[f_{vec}]=\frac{1}{2}(1-e^{2ia})(1-e^{-2ia})\\\nn
&&\prod_{i,j=1}^{\infty}\Big[\Big(1-x^{i+j-1}y^{i-j+1}\Big)
\Big(1-x^{i+j-1}y^{i-j-1}\Big)
\Big(1-e^{2ia}\,x^{i+j-1}y^{i-j+1}\Big)\times\\\nn
&&\Big(1-e^{2ia}\,x^{i+j-1}y^{i-j-1}\Big)\Big(1-e^{-2ia}\,x^{i+j-1}y^{i-j+1}\Big)\Big(1-e^{-2ia}\,x^{i+j-1}y^{i-j-1}\Big)\Big]\\\nn
&=&\frac{1}{2}\Big|(1-e^{2ia})\prod_{i,j=1}^{\infty}\Big(1-x^{i+j-1}y^{i-j+1}\Big)\Big(1-e^{2ia}\,x^{i+j-1}y^{i-j+1}\Big)\Big(1-e^{2ia}\,x^{i+j-1}y^{i-j-1}\Big)\Big|^2
\eea

\noindent Comparing Eq.(\ref{g1}), Eq.(\ref{g2}), Eq.(\ref{g3}) and Eq.(\ref{indexpert}) we see that
\bea
2\,\mbox{sin}^{2}(\alpha)PE[f_{vec}]=\frac{1}{2}\Big|M(t,q)\,Z_{0}(Q_{f},t,q)\Big|^{2}
\eea
Thus the perturbative part of the integrand in Eq.(\ref{kim}) is exactly given by the part of the topological string partition function which gets contributions from the D0-branes and D2-branes wrapping the fiber curve. The instanton part of the integrand in Eq.(\ref{kim}) is precisely the Nekrasov's instanton partition function. It is known that Nekrasov's instanton partition function for $SU(2)$ with $N_{f}=0$ is precisely equal to the part of the topological string partition function which includes contributions from the base curve i.e., $Z'(Q_{b},Q_{f},t,q)$ given by Eq.(\ref{nekins}).

\subsection{Example 2: Blowup of local $\mathbb{P}^{1}\times \mathbb{P}^{1}$}
\begin{figure}[h]
  \centering
  \includegraphics[width=3in]{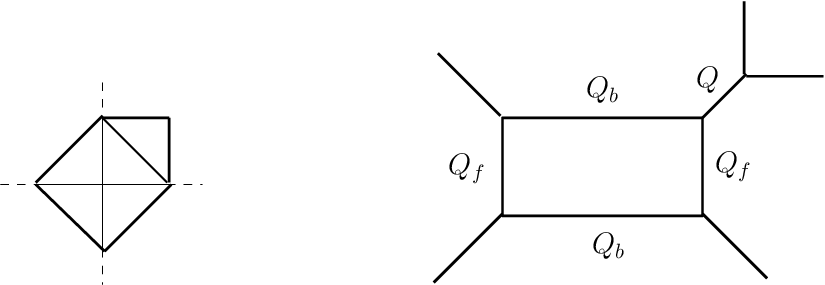}\\
  \caption{}\label{blowupp1p1}
\end{figure}

The blowup of local $\mathbb{P}^{1}\times \mathbb{P}^{1}$ is another interesting example that we will work out in this section. The Newton polygon and the web diagram of this geometry is shown in \figref{blowupp1p1} below

The $H_{2}(X,\mathbb{Z})$ is spanned by $\{B,F,E\}$ where $B$ and $F$ are base and the fiber curves and $E$ is the exceptional curve coming from blowup. The intersection numbers are given by
\bea\label{inter2a}
B\cdot B=0\,,\,\,F\cdot F=0\,,\,\,B\cdot F=+1\,,\,\,B\cdot E=F\cdot E=0\,,\,\,E\cdot E=-1\,.
\eea
The anticanonical class is given by
\bea
-K_{X}=2(B+F)-E
\eea
and using Eq.(\ref{inter2a}) we get
\bea
-K_{X}\cdot B=+2\,,\,\,\,-K_{X}\cdot F=+2\,,\,\,-K_{X}\cdot E=+1\,.
\eea
As discussed before the intersection number of the curves with the aniticanonical class (the degree of the curve) determines the electric charges of the state coming from M2-brane wrapping the curve and determines the relation between the loop variables and the K\"ahler parameters. In this case we get\footnote{As before we have chosen the position of the external line of the web diagram such that $t_{f}\mapsto 0$ as $a\mapsto 0$.}
\bea\label{abc}
Q_{b}=u\,e^{2ia}\,,\,\,Q_{f}=e^{2ia}\,,\,\,\,Q=\tilde{u}e^{-ia}\,.
\eea

The refined partition function of this geometry is given by
\bea
Z_{X}(Q_{b},Q_{f},Q,t,q)&=&\Big(M(t,q)M(q,t)\Big)^{\frac{1}{2}}\,Z(Q_{b},Q_{f},t,q)\\\nn
Z(Q_{b},Q_{f},t,q)&:=&\sum_{\nu_{1}\,\nu_{2}}Q_{b}^{|\nu_{1}|+|\nu_{2}|}\,\,q^{||\nu_{2}^{t}||^2}\,t^{||\nu_{1}^{t}||^2}\widetilde{Z}_{\nu_{1}}(t,q)\widetilde{Z}_{\nu_{2}}(q,t)
\widetilde{Z}_{\nu_{1}^t}(q,t)\widetilde{Z}_{\nu_{2}^t}(t,q)\times \\\nn
&&\prod_{i,j=1}^{\infty}
\frac{\Big(1-Q\,t^{i-\frac{1}{2}-\nu_{1,j}^t}q^{j-\frac{1}{2}}\Big)
\Big(1-Q\,Q_{f}t^{i-\frac{1}{2}-\nu_{1,j}}q^{j-\frac{1}{2}}\Big)}{\Big(1-Q_{f}t^{i-1-\nu_{2,j}}q^{j-\nu_{1,i}}\Big)\Big(1-Q_{f}q^{i-1-\nu_{1,j}}t^{j-\nu_{2,i}}\Big)}\,.
\eea
Using the refined partition function and Eq.(\ref{abc}) the index of this geometry is given by
\bea
I_{X}(u,\tilde{u},t,q)&=&\int da\,\Big|Z_{X}(u\,e^{2ia},e^{2ia},\tilde{u}e^{ia},t,q)\Big|^2\,\\\nn
&=&1+\Big(2+\tilde{u}+\frac{1}{\tilde{u}}\Big)x^2+\Big(3y+\frac{3}{y}+\tilde{u}y+\frac{\tilde{u}}{y}+\frac{1}{y\,\tilde{u}}+\frac{y}{\tilde{u}}\Big)x^3+\cdots
\eea

This agrees with the result of \cite{KKL}.

A more detailed analysis can be carried out in this case to identify different pieces of the gauge theoretic calculation and the topological string calculation. The gauge theory calculation of \cite{KKL} gives the index to be
\bea\nn
I=\int da \underbrace{2\mbox{sin}^{2}(a)\mbox{PE}\Big[f_{vec}(a,x,y)+f_{matter}(a,m,x,y)\Big]}_{perturbative\,\,contribution}\,\underbrace{\Big|Z_{instanton}(a,q,m,x,y)\Big|^2}_{instanton\,\,contribution}\,.
\eea
In the previous example we have already shown that part of the above perturbative contribution that depends on the Haar measure and $f_{vec}(a,x,y)$ comes from fiber curve and the D0-brane contribution (the constant map configurations in the worldsheet terms). The new contribution to the perturbative part here is the term that depends on $f_{matter}(a,m,x,y)$ where $(x=\sqrt{\frac{q}{t}},y=\sqrt{q\,t})$
\bea\label{fm}
f_{matter}(a,x,y,m)&=&\frac{x}{(1-xy)(1-\frac{x}{y})}
(e^{-ia-im}+e^{ia-im}+
e^{-ia+im}+e^{ia+im})\\\nn
&=&-\frac{1}{(q^{\frac{1}{2}}-q^{-\frac{1}{2}})(t^{\frac{1}{2}}-t^{-\frac{1}{2}})}(e^{-ia-im}+e^{ia-im}+
e^{-ia+im}+e^{ia+im})\,.
\eea

\noindent It is easy to see that the contribution of this term to the perturbative part obtained through the plethystic exponential is precisely equal to the contribution of the curve $E$ and $F+E$ to the partition function and its complex conjugate. These are the only holomorphic curves that do not involve the curve $B$ (which would be the instanton contribution). Since the curve $E$ and $F+E$ are locally both $(-1,-1)$ curves therefore they are rigid and have $N^{j_{L},j_{R}}_{E}=N^{j_{L},j_{R}}_{E+F}=\delta_{j_{L},0}\delta_{j_{R},0}$ and therefore from Eq.(\ref{rpf}) the contribution to the partition function from these curves is given by
\bea
\widetilde{Z}(Q,Q_{f},t,q)=\prod_{i,j}\Big(1-Q\,q^{i-\frac{1}{2}}\,t^{j-\frac{1}{2}}\Big)
\Big(1-Q\,Q_{f}q^{i-\frac{1}{2}}\,t^{j-\frac{1}{2}}\Big)\,.
\eea
From Eq.(\ref{fm}), the definition of the pleythestic exponential and the above equation it follows that
\bea\nn
PE[f_{matter}(a,m,x,y)]&=&\prod_{i,j=1}^{\infty}
\Big(1-e^{-ia-im_{\ell}}q^{i-\frac{1}{2}}\,t^{j-\frac{1}{2}}\Big)
\Big(1-e^{ia-im_{\ell}}q^{i-\frac{1}{2}}\,t^{j-\frac{1}{2}}\Big)\times\\\nn
&&\Big(1-e^{-ia+im_{\ell}}q^{i-\frac{1}{2}}\,t^{j-\frac{1}{2}}\Big)
\times\Big(1-e^{ia+im_{\ell}}q^{i-\frac{1}{2}}\,t^{j-\frac{1}{2}}\Big)\\\nn
&=&\widetilde{Z}(Q,Q_{f},t,q)\widetilde{Z}(Q^{-1},Q_{f}^{-1},t,q)=\Big|\widetilde{Z}(Q,Q_{f},t,q)\Big|^2\,,\\\nn
\eea
where
\bea
Q=e^{-ia-im}\,,\,\,\,\,Q\,Q_{f}=e^{ia-im}\,.
\eea

\subsection{Example 3: Local $\mathbb{F}_{1}$}

\begin{figure}[h]
  \centering
  \includegraphics[width=4in]{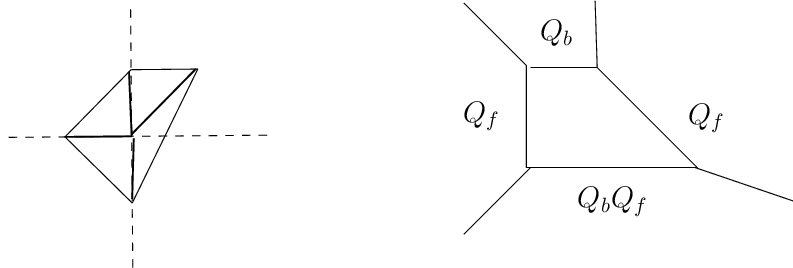}\\
  \caption{The newton polygon (a) and the  web diagram (b) of local $\mathbb{F}_{1}$. In this case there are two distinct triangulations of the Newton polygon corresponding to two different phases.}\label{localf1}
\end{figure}

Let us consider the CY threefold which is the total space of canonical bundle on the Hirzebruch surface $\mathbb{F}_{1}$. $\mathbb{F}_{1}$ is a non-trivial $\mathbb{P}^{1}$ bundle over $\mathbb{P}^{1}$, we will denote the base $\mathbb{P}^{1}$ by $B$ and the fiber $\mathbb{P}^{1}$ by $F$ with corresponding K\"ahler parameter $t_{b}$ and $t_{f}$ respectively, such that $B\cdot B=-1$. As usual we define $Q_{b}=e^{-t_{b}}$ and $Q_{f}=e^{-t_{f}}$. The Newton polygon and the web diagram of local $\mathbb{F}_{1}$ is shown in \figref{localf1}.

\noindent The refined partition function for this geometry was calculated in \cite{IKV} and is given by (see Appendix A for notation and other details)
\bea\label{f1pf}
Z_{\tiny local\, \mathbb{F}_{1}}(Q_{b},Q_{f},t,q)&:=&(M(t,q)M(q,t)^{\frac{1}{2}}\,Z(Q_{b},Q_{f},t,q)\\\nn
Z(Q_{b},Q_{f},t,q)&=&\sum_{\nu_{1},\nu_{2}}Q_{b}^{|\nu_{1}|+|\nu_{2}|}Q_{f}^{|\nu_{2}|}(-1)^{(|\nu_{1}|+|\nu_{2}|)}
\Big(\frac{q}{t}\Big)^{\frac{||\nu_{1}||^2+||\nu_{2}||^2}{2}}
t^{\frac{\kappa(\nu_{1})-\kappa(\nu_{2})}{2}}\\\nn
&&\,q^{||\nu_{2}^t||^2}\,t^{||\nu_{1}^t||^2}\,\widetilde{Z}_{\nu_{1}}(t,q)\widetilde{Z}_{\nu_{1}^t}(q,t)\widetilde{Z}_{\nu_{2}}(q,t)\widetilde{Z}_{\nu_{2}^t}(t,q)\times\\\nn
&&
\prod_{i,j=1}^{\infty}\Big[(1-Q_{f}t^{i-\nu_{2,j}}q^{j-1-\nu_{1,i}})
(1-Q_{f}q^{i-\nu_{1,j}}t^{j-1-\nu_{2,i}})\Big]^{-1}
\eea

\begin{figure}[h]
  \centering
  \includegraphics[width=5in]{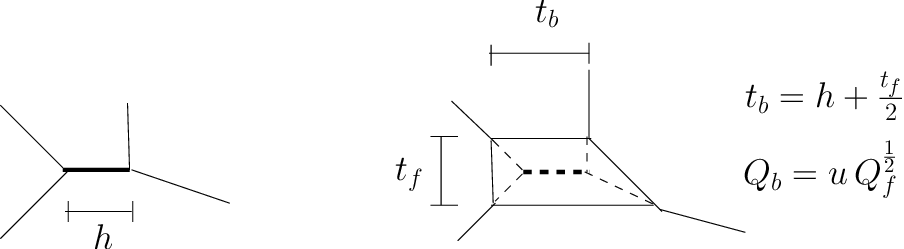}\\
  \caption{}\label{loopf1}
\end{figure}

\noindent The index for this geometry is therefore given by
\bea
I_{local \,\mathbb{F}_{1}}=\int da\,\Big|Z_{\tiny local\, \mathbb{F}_{1}}(Q_{b},Q_{f},t,q)\Big|^{2}\,,
\eea
where $a$ is the loop variable corresponding to the only 4-cycle in the geometry. In order to calculate the index we have to determine the dependence of the K\"ahler parameters $t_{b}$ and $t_{f}$ on the loop variable. It is easy to see from the general result given in section 6 that
\bea
Q_{f}=e^{2ia}
\eea
The geometry of the web determines the relation between $t_{f}$ and $t_{b}$. If we fix the external legs of the web we can change the size of the 4-cycle by changing $t_{f}$, if we take $t_{f}=0$ then the web diagram is shown in \figref{loopf1} and the parameter $h$ is determined by the position of the external legs. The index will be a function of this parameter $h$ (along with $x$ and $y$). The relation between the $Q_{b}$ and $Q_{f}$ can be easily determined from the web diagram and is given by
\bea
Q_{b}=u\,Q_{f}^{\frac{1}{2}}
\eea

\noindent Thus the index is given by
\bea\label{indexf1}
I_{\tiny local \,\mathbb{F}_{1}}&=&\int da\,\Big|Z_{\tiny local\, \mathbb{F}_{1}}(u\,e^{ia},e^{2ia},t,q)\Big|^{2}\\\nn
&=&\int \frac{1}{2}\frac{dQ_{f}}{2\pi i\,Q_{f}}\,\Big|Z_{\tiny local\, \mathbb{F}_{1}}(u\,Q_{f}^{\frac{1}{2}},Q_{f},t,q)\Big|^{2}\,.
\eea

\noindent Using Eq.(\ref{f1pf}) and Eq.(\ref{indexf1}) we get
\bea\nn
I_{\tiny local \mathbb{F}_{1}}&=&1+x^{2}+2\Big(y+\frac{1}{y}\Big)x^{3}+\Big(3+2y^{2}+\frac{2}{y^2}\Big)x^4+
\Big(2y^{3}+3y+\frac{3}{y}+\frac{2}{y^3}\Big)x^5+\\\nn&&
\Big(u^{2}+\frac{1}{u^2}+5+2y^{4}+5y^{2}+
\frac{5}{y^2}+\frac{2}{y^4}\Big)x^6+\\\nn&&\Big(2y^{5}+6y^{3}+u^{2}y+10y+\frac{y}{u^{2}}+\frac{u^{2}}{y}+
\frac{1}{u^{2}y}+\frac{10}{y}
+\frac{6}{y^{3}}+\frac{2}{y^{5}}\Big)x^{7}+\cdots
\eea

\subsubsection{The flop invariance of the index}\label{fi}
Recall that the Newton polygon of the local $\mathbb{F}_{1}$ has two distinct triangulations. These two triangulations correspond to two different geometries which are related with each other by a flop transition. Here we will show that the index we have computed above is invariant under the flop transition.

\begin{figure}[h]
  \centering
  \includegraphics[width=4in]{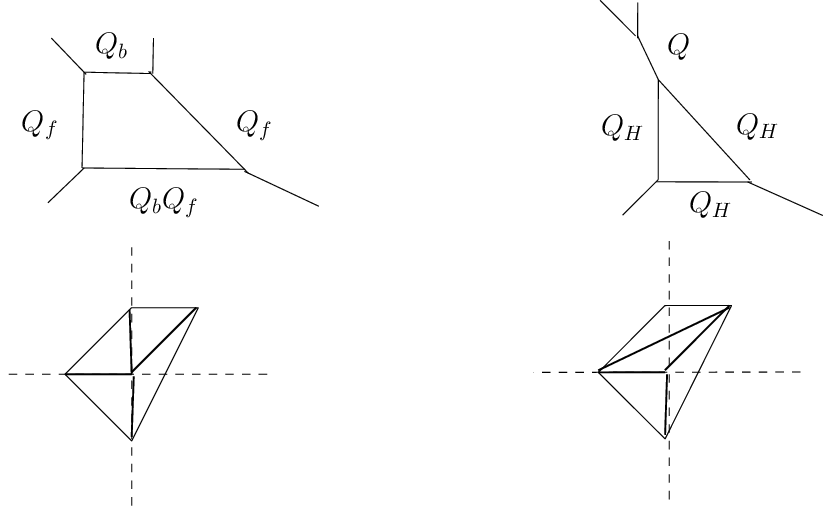}\\
  \caption{}\label{flop}
\end{figure}

\noindent In \figref{flop} the two triangulations and the corresponding web diagrams are shown. The the neighborhood of the base curve $B$ of the local $\mathbb{F}_{1}$ is ${\cal O}(-1)\oplus {\cal O}(-1)\mapsto \mathbb{P}^{1}$ which can undergo flop giving the resulting geometry which is local $\mathbb{P}^{2}$ together with the flopped curve. We will denote by $X$ the local $\mathbb{F}_{1}$ geometry and will denote by $Y$ the geometry obtained by flop from $X$.\\

\noindent The relation between the K\"ahler parameters on the two sides of the flop is given by
\bea
Q_{H}=Q_{b}Q_{f}\,,\,\,\,\,Q=Q_{b}^{-1}\,.
\eea
\noindent The refined partition function of $Y$ can be determined from the partition function of $X$ by carefully following flop and is given by \cite{paper}
\bea\label{rel}
Z_{Y}(Q_{H},Q,t,q)=Z_{\tiny local\, \mathbb{F}_{1}}(Q^{-1},Q_{H}Q,t,q)\,.
\eea

\begin{figure}[h]
  \centering
  \includegraphics[width=3in]{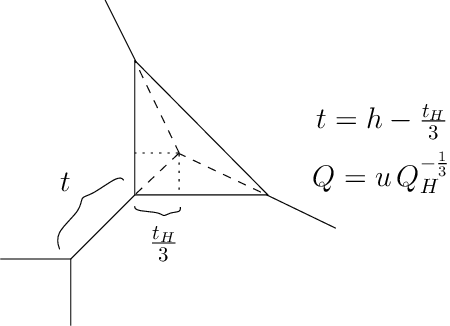}\\
  \caption{}\label{loopflop}
\end{figure}

\noindent In the corresponding web diagram \figref{flop}(b) if we fix the external legs then the size of the $\mathbb{P}^{1}$, which came from the flop, changes with the size of the $\mathbb{P}^{2}$ and the relation between them, as shown in \figref{loopflop}, is given by:
\bea
Q=u\,Q_{H}^{-\frac{1}{3}}\,,\,\,\,\,u=e^{-h}\,,
\eea
where $h$ is the size of the flopped curve when $t_{H}=0$.

\noindent The index of $Y$ is then given by
\bea\label{indexflop}
I_{Y}(u,t,q):=\int da\Big|Z_{Y}(Q_{H},u\,Q_{H}^{-\frac{1}{3}},t,q)\Big|^{2}\,,
\eea
where $a$ is the loop variable corresponding to the 4-cycle $\mathbb{P}^{2}$. The relation between the loop variable $a$ and the K\"ahler parameter $t_{H}$ is given by
\bea
t_{H}=3a\,.
\eea
Eq.(\ref{indexflop}) becomes
\bea
I_{Y}(u,t,q)=\int \frac{dz}{2\pi iz} \Big|Z_{Y}(z^3,u\,z^{-1},t,q)\Big|^{2}\,.
\eea
where $z=e^{ia}$. Using the relation between the partition function of $Y$ and that of $X$ Eq.(\ref{rel}) then gives
\bea
I_{Y}(u,t,q)=\int \frac{dz}{2\pi iz}\Big|Z_{X}(u^{-1}\,z,u\,z^2,t,q)\Big|^2\,.
\eea
Changing the integration variable $z\mapsto \frac{z}{\sqrt{u}}$ we get
\bea
I_{Y}(u,t,q)&=&\int \frac{dz}{2\pi i\,z}\Big|Z_{X}(u^{-\frac{3}{2}}\,z,z^2,t,q)\Big|^2\,,\\\nn
&=&\int da\,\Big|Z_{X}(u^{-\frac{3}{2}}\,e^{i\,a},e^{2i\,a},t,q)\Big|^2\,.
\eea
Comparing the above with Eq.(\ref{indexf1}) we see that
\bea
I_{Y}(u,t,q)=I_{X}(u^{-2/3},t,q)
\eea

\subsection{Example 4: Local $\mathbb{F}_{2}$}

\begin{figure}[h]
  \centering
  \includegraphics[width=4in]{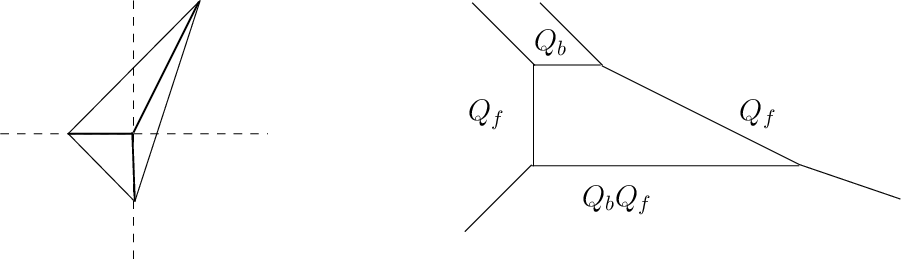}\\
  \caption{The Newton polygon (a) and the  web diagram (b) of local $\mathbb{F}_{2}$.}\label{localf2}
\end{figure}

The Hirzebruch surface $\mathbb{F}_{2}$ is also a $\mathbb{P}^{1}$ bundle over $\mathbb{P}^{1}$. The total space of the canonical bundle on $\mathbb{F}_{2}$ gives a local CY threefold. As before the K\"ahler parameters corresponding to the base $B$ and the fiber $F$ will be called $t_{b}$ and $t_{f}$ and we define $Q_{b}:=e^{-t_{b}}$ and $Q_{f}:=e^{-t_{f}}$. The Newton polygon and the web diagram of local $\mathbb{F}_{2}$ is shown in \figref{localf2}.

\noindent The refined partition function for this geometry is given by
\bea\label{f2pf}
Z_{\tiny local\, \mathbb{F}_{2}}(Q_{b},Q_{f},t,q)&=&(M(t,q)M(q,t)^{\frac{1}{2}}\,Z(Q_{b},Q_{f},t,q)\\\nn
Z(Q_{b},Q_{f},t,q)&=&\sum_{\nu_{1},\nu_{2}}Q_{b}^{|\nu_{1}|+|\nu_{2}|}Q_{f}^{2|\nu_{2}|}
\Big(\frac{q}{t}\Big)^{||\nu_{1}||^2+||\nu_{2}||^2}
t^{\kappa(\nu_{1})-\kappa(\nu_{2})}\,q^{||\nu_{2}^t||^2}\,t^{||\nu_{1}^t||^2}\\\nn
&&\,\widetilde{Z}_{\nu_{1}}(t,q)\widetilde{Z}_{\nu_{1}^t}(q,t)\widetilde{Z}_{\nu_{2}}(q,t)\widetilde{Z}_{\nu_{2}^t}(t,q)\times\\\nn
&&
\prod_{i,j=1}^{\infty}\Big[(1-Q_{f}t^{i-\nu_{2,j}}q^{j-1-\nu_{1,i}})
(1-Q_{f}q^{i-\nu_{1,j}}t^{j-1-\nu_{2,i}})\Big]^{-1}
\eea

\begin{figure}[h]
  \centering
  \includegraphics[width=5in]{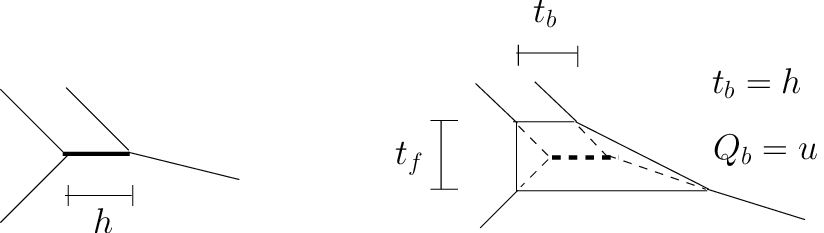}\\
  \caption{In this case since the there are two parallel legs the area of the base curve does not change with the area of the fiber curve.}\label{loopf2}
\end{figure}

\noindent The index is given by
\bea
I_{\tiny local\, \mathbb{F}_{2}}=\int da \,\Big|Z_{\tiny local\, \mathbb{F}_{2}}(Q_{b},Q_{f},t,q)\Big|^{2}
\eea
The relation between the loop variable $a$ and the fiber parameter $t_{f}$ is the same as before
\bea
t_{f}=2a\,.
\eea
Just as before we fix the external legs of the web so that we have one parameter $h$ as shown in \figref{loopf2}. The relation between the $Q_{b}$ and $Q_{f}$ in this case is given by \footnote{In general for local $\mathbb{F}_{m}$ the relation is $Q_{b}=u\,Q_{f}^{1-\frac{m}{2}}$.}
\bea
Q_{b}=e^{-h}=u\,
\eea

\noindent Thus the index becomes
\bea
I_{local F_{2}}=\int da\,\Big|Z_{\tiny local\, \mathbb{F}_{2}}(u\,,e^{2ia}\,,t\,,q)\Big|^2
\eea

\noindent Using Eq.(\ref{f2pf}) we get
\bea\nn
I_{local\,\mathbb{F}_{2}}&=&1+x^{2}+2\Big(y+\frac{1}{y}\Big)x^{3}+\Big(3+2y^{2}+\frac{2}{y^2}\Big)x^4+\\\nn
&&
\Big(2y^{3}+3y+\frac{3}{y}+\frac{2}{y^3}-(u+\frac{1}{y})(y+\frac{1}{y})\Big)x^5\\\nn&&
\Big(2y^{4}+5y^{2}+\frac{5}{y^{2}}+\frac{2}{y^4}+4-(u+\frac{1}{u})(3+y^2+\frac{1}{y^2})\Big)x^6+
\\\nn&&\Big(2y^{5}+6y^{3}+9y+\frac{9}{y}+\frac{6}{y^3}+\frac{2}{y^5}-(u+\frac{1}{u})(y^{3}+3y+\frac{3}{y}+\frac{1}{y^3})\Big)x^{7}+\cdots
\eea

\subsection{Example 5: Local $\mathbb{P}^{2}$}

\begin{figure}[h]
  \centering
  \includegraphics[width=1.75in]{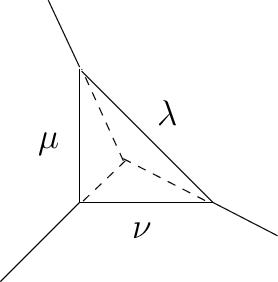}\\
  \caption{The web diagram of local $\mathbb{P}^2$.}\label{P2}
\end{figure}

The local $\mathbb{P}^{2}$ is the total space of ${\cal O}(-3) \mapsto \mathbb{P}^{2}$. As discussed in section 4 the refined topological vertex alone can not be used to calculate its partition function since there is no set of edges which cover the vertices and are parallel to each other. However, some recent developments have made it possible to calculate the refined partition function for any local toric CY3fold \cite{indexvertex,AS2,paper}. We use the form of the partition function given in \cite{paper}. The web diagram of the local $\mathbb{P}^{2}$ is shown in \figref{P2}.

\noindent The refined partition function of local $\mathbb{P}^{2}$ is given by \cite{paper, AS2}
\bea
Z_{local \,\mathbb{P}^2}(Q,t,q)&=&\sum_{\lambda\,\mu\,\nu}(-Q)^{|\lambda|+|\mu|+|\nu|}\,q^{\frac{3||\nu^t||^2}{2}}\,t^{-\frac{||\nu||^2}{2}}\,\widetilde{Z}_{\nu}(q,t)
 \widetilde{Z}_{\nu^t}(t,q)\times\\\nn&&s_{\lambda}(q^{-\rho}t^{-\nu})\,s_{\mu}(q^{-\rho}t^{-\nu})\,\Big(\frac{q}{t}\Big)^{\frac{|\lambda|-|\mu|}{2}}\,N^{\eta}_{\lambda\,\mu}
 R_{\eta}\,,
\eea
where
\bea\nn
R_{\eta}=\sum_{\sigma}\,U_{\eta\,\sigma}\,t^{\frac{||\sigma^t||^2}{2}}\,q^{-\frac{||\nu||^2}{2}}\,P_{\sigma}(t^{-\rho};q,t)\,.
\eea
$N^{\eta}_{\lambda\,\mu}$ are the Littlewood-Richardson coefficients and $U_{\eta\sigma}$ is the matrix which takes Macdonald polynomials to Schur polynomials, $s_{\eta}({\bf x})=\sum_{\sigma}U_{\eta\,\sigma}\,P_{\sigma}({\bf x};q,t)$. The matrix elements $U_{\eta\,\sigma}$ are rational functions of $q$ and $t$, for example:
\bea\nn
U_{(1)\,(1)\,}&=&1\\\nn
U_{(2)\,(2)}&=&1\,,\,\,\,U_{(2)\,(1\,1)}=\frac{t-q}{1-t\,q}\,,\,\,\,U_{(1\,1)\,(2)}=0\,,\,\,\,\,U_{(1\,1)\,(1\,1)}=1\,.
\eea

\noindent The above partition function can also be written as
\bea\label{exp}
Z(Q,t,q)&=&\sum_{\nu}(-Q)^{|\nu|}\,\Big[q^{\frac{3||\nu^t||^2}{2}}t^{-\frac{||\nu||^2}{2}}\,
\widetilde{Z}_{\nu}(q,t)\widetilde{Z}_{\nu^t}(t,q)\Big]
Z_{\nu}(Q,t,q)\,,\eea
where\bea\label{fibersum}
Z_{\nu}(Q,t,q)&=&\sum_{\lambda\,\mu\,\eta}(-Q)^{|\lambda|+|\mu|}s_{\lambda}(q^{-\rho}\,t^{-\nu})s_{\nu}(q^{-\rho}t^{-\nu})\Big(\frac{q}{t}\Big)^{\frac{|\lambda|-|\mu|}{2}}\,N^{\eta}_{\lambda\,\mu}
 R_{\eta}\\\nn
\eea
In Eq.(\ref{exp}) the factor in the square bracket has expansion in positive powers of $q$ and $t^{-1}$ (i.e., positive powers of $x$) but the factor in the second line, $Z_{\nu}$, has expansion in positive powers of $q^{-1}$ and $t$ (i.e., negative powers of $x$). This is the generic case for partition functions calculated using the refined topological vertex and can be understood from Eq.(\ref{pe}). Since the variables $x$ and $y$ couple to the $SU(2)_{R}$ and $SU(2)_{L}$ therefore as long as we have full $(j_{L},j_{R})$ spin content negative power of $x$ can not be avoided. However, in certain special cases we can sum over a class of curves and get a product representation of the a part of the partition function which allows up to expand the partition function in positive powers of $x$ at the expense of introducing $Q$ and $Q^{-1}$ in the expansion. This is how the index as expansion inn positive powers of $x$ was determined in the last three examples. Therefore what is required here is some way of summing up the contribution from the curves labelled by $\lambda$ and $\mu$ in \figref{P2}, $Z_{\nu}(Q,t,q)$, to obtain a product representation which can then be expanded in positive powers of $x$.\\

\noindent The index is then given by
\bea
I_{\tiny local\,\mathbb{P}^{2}}=\int da\,\Big|Z_{local \,\mathbb{P}^2}(Q,t,q)\Big|^2
\eea
The relation between the loop variable and $Q$ is the same as we derived in showing the flop invariance of the index in Section \ref{fi},
\bea
Q=e^{3ia}\,.
\eea
Thus the index becomes
\bea
I_{\tiny local\,\mathbb{P}^{2}}=\int da\,\Big|Z_{local \,\mathbb{P}^2}(e^{3ia},t,q)\Big|^2
\eea

\subsection{Example 6: Flop invariance of the index}
\begin{figure}[h]
  \centering
  \includegraphics[width=4in]{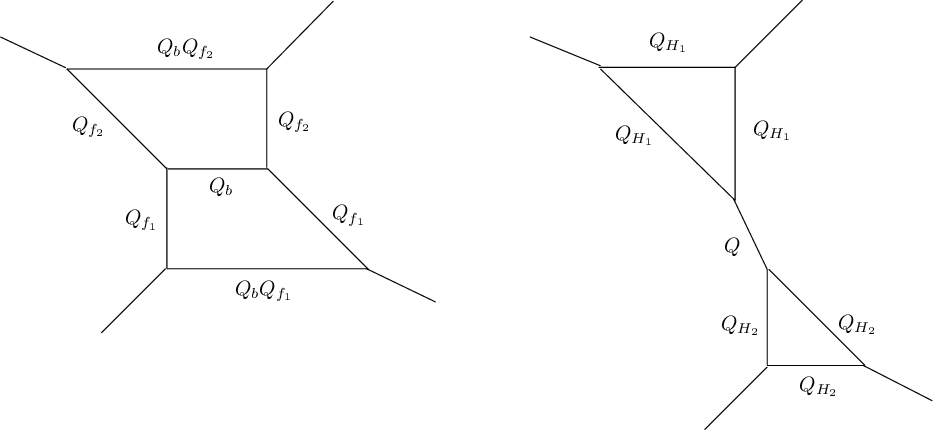}\\
  \caption{}\label{flop2}
\end{figure}

Here we present another example which shows that the index is invariant under flop transition. The web diagram of the geometry we will discuss is shown in \figref{flop2}. The geometry consists of two 4-cycles $D_{1}$ and $D_{2}$, both Hirzebruch surface $\mathbb{F}_{1}$, intersecting along a $\mathbb{P}^{1}$ which is the base of the fibration for both divisors. In the neighbourhood of the base curve the geometry looks like ${\cal O}(-1)\oplus {\cal O}(-1)\mapsto \mathbb{P}^{1}$ and therefore the base curve can flopped as shown in \figref{flop2}.

\noindent We will call the geometry before the flop (the one with two $\mathbb{F}_{1}$ divisors) $X$ and the geometry after the flop (the one with two $\mathbb{P}^{2}$ divisors) $Y$. The relation between the K\"ahler parameters of geometry $X$ and $Y$ is given by
\bea
Q_{H_{1}}=Q_{b}Q_{f_{1}}\,,\,\,Q_{H_{2}}=Q_{b}Q_{f_{2}}\,,\,\,\,Q=Q_{b}^{-1}\,.
\eea
The relation between the partition functions follows from the above relation between the K\"ahler parameters,
\bea\label{rel2}
Z_{Y}(Q_{H_{1}},Q_{H_{2}},Q,t,q)=Z_{X}(Q^{-1},Q_{H_{1}}Q\,,Q_{H_{2}}Q,t,q)
\eea

\noindent The geometry $X$ has two 4-cycles and hence two loop variables $a_{1}$ and $a_{2}$. As discussed in section 5 the relation between the K\"ahler parameters and the loop variables can be determined using the intersection between curves and the 4-cycles i.e., intersection of the curve and the anticanonical class of the 4-cycle. Let us denote the 4-cycles in $X$ by $D_{1}$ and $D_{2}$ then the anticanonical class is given by
\bea
-K_{D_{1}}=2B+3F_{1}\,,\,\,\,\,-K_{D_{2}}=2B+3F_{2}\,.
\eea.
 Then the charge vector of the curve $C=n\,B=m_{1}F_{1}+m_{2}F_{2}$ is given by
 \bea\label{inter}
 \vec{d}=(-K_{D_{1}}\cdot C,\,-K_{D_{2}}\cdot C)=(n+2m_{1}-m_{1},n-m_{1}+2m_{2})\,,
 \eea
 where we have used the following intersection numbers in calculating Eq.(\ref{inter})
 \bea
 B\cdot B=-1\,,\,\,B\cdot F_{1}=+1\,,\,\,B\cdot F_{2}=+1\,,\,\,F_{1}\cdot F_{2}=-1\,.
 \eea
 Thus the K\"ahler parameter corresponding to $C$ in terms of the loop variables is given by
 \bea
 Q_{C}=Q_{C,0}e^{i\vec{d}\cdot \vec{a}}=e^{i(n+2m_{1}-m_{2})a_{1}+i(n-m_{1}+2m_{2})a_{2}}
 \eea
 Thus for geometry $X$
\bea
Q_{f_{1}}=e^{i(2a_{1}-a_{2})}\,,\,Q_{f_{2}}=e^{i(2a_{2}-a_{1})}\,,\,Q_{b}=u\,e^{i(a_{1}+a_{2})}\\\nn
\eea

In the case geometry $Y$ the two 4-cycles will be denoted by $P_{1}$ and $P_{2}$ with corresponding loop variables $b_{1}$ and $b_{2}$ respectively. Both these divisors are $\mathbb{P}^2$ and the anticanonical class of these divisors is given by
\bea
-K_{P_{1}}=3H_{1}-E\,,\,\,\,\,-K_{P_{2}}=3H_{2}-E\,,
\eea
where $H_{1}$ is the hyperplane class of $P_{1}$, $H_{2}$ is the hyperplane class of $P_{2}$ and $E$ is the curve connecting the two which comes from the flop of the curve $B$. The intersection numbers of these curves are
\bea
H_{1}\cdot H_{1}=1\,,\,\,H_{2}\cdot H_{2}=1\,,\,\,H_{1}\cdot H_{2}=0\,,\,\,H_{1}\cdot E=H_{2}\cdot E=0\,.
\eea
Using the above intersection numbers we can easily determine the charge vector of the curve $C=nH_{1}+mH_{2}-kE$,
\bea
\vec{d}=(-K_{P_{1}}\cdot C,-K_{P_{2}}\cdot C)=(3n-k,3m-k)\,,
\eea
thus the K\"ahler parameter of $C$ scales with loop variables as
\bea
Q_{C}=Q_{C,0}\,e^{i\vec{d}\cdot\vec{b}}=Q_{C,0}e^{i(3n-k)b_{1}+(3m-k)b_{2}}\,.
\eea
Thus for geometry $Y$
\bea
Q_{H_{1}}=e^{3ib_{1}}\,,\,\,Q_{H_{2}}=e^{3ib_{2}}\,,\,\,Q=Q_{E}=\tilde{u}\,e^{-i(b_{1}+b_{2})}
\eea
Now that we have the relation between the K\"ahler parameters and the loop variables we can discuss the index of the two geometries. The index for geometry $X$ and $Y$ is given by
\bea\nn
I_{X}(u,t,q)&=&\int da_{1}da_{2}\,\Big|Z_{X}(Q_{b},Q_{f_{1}},Q_{f_{2}})\Big|^2\\\nn
&=&\int da_{1}da_{2}\,\Big|Z_{X}(ue^{i(a_{1}+a_{2})},e^{i(2a_{1}-a_{2})},e^{i(2a_{2}-a_{1})})\Big|^2\\\nn
I_{Y}(\tilde{u},t,q)&=&\int db_{1}db_{2}\,\Big|Z_{Y}(Q_{H_{1}},Q_{H_{2}},Q)\Big|^2\\\nn
&=&\int\,db_{1}db_{2}\,
\Big|Z_{Y}(e^{3ib_{1}},e^{3ib_{2}},\tilde{u}e^{-i(b_{1}+b_{2})})\Big|^2\,.\\
\eea
Now using Eq.(\ref{rel2}) we get
\bea
I_{Y}(\tilde{u},t,q)&=&\int db_{1}\,db_{2}\,\Big|Z_{X}(\tilde{u}^{-1}\,e^{i(b_{1}+b_{2})},\tilde{u}e^{i(2b_{1}-b_{2})}\,\tilde{u}e^{i(2b_{2}-b_{1})}\Big|^2\\\nn
&=&\int \frac{dz_{1}}{2\pi i\,z_{1}}\,\frac{dz_{2}}{2\pi i\, z_{2}}
\Big|Z_{X}(\tilde{u}^{-1}z_{1}z_{2},\tilde{u}z_{1}^2\,z_{2}^{-1}\,,\tilde{u}z_{2}^{2}\,z_{1}^{-1}\Big|^2\\\nn
\eea
Let $z_{1}\mapsto \tilde{u}^{-1}\,z_{1},z_{2}\mapsto \tilde{u}^{-1}\,z_{2}$ then we get
\bea
I_{Y}(\tilde{u},t,q)&=&\int \frac{dz_{1}}{2\pi i\,z_{1}}\,\frac{dz_{2}}{2\pi i\, z_{2}}
\Big|Z_{X}(\tilde{u}^{-3}z_{1}z_{2},\,z_{1}^2\,z_{2}^{-1}\,,\,z_{2}^{2}\,z_{1}^{-1}\Big|^2\\\nn
&=&I_{X}(\tilde{u}^{-3},t,q)\,.
\eea
Which proves the flop invariance of the index.

\subsection{Computation of the index with 3d defects}
As discussed in section 3 3D defects in the 5D theory can be engineered using Lagrangian branes. In this section we consider some examples in which there is a single Lagrangian brane in the geometry.

\subsubsection{Lagrangian brane on $\mathbb{C}^3$}
Let us begin by considering the simplest of the brane configurations, a Lagrangian brane on $\mathbb{C}^3$. The geometry is shown in \figref{c3} below.

\begin{figure}[h]
  \centering
  \includegraphics[width=1in]{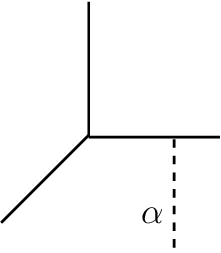}\\
  \caption{}\label{c3}
\end{figure}

The partition function of the brane is given
\bea
Z_{Brane}(Q,U,t,q)&=&\sum_{\alpha}\,(-Q)^{|\alpha|}\,\mbox{Tr}_{\alpha}U\,\,s_{\lambda^t}(q^{-\rho})\,,\\\nn
\eea
where $U$ is the holonomy on the brane and $-\mbox{log}Q$ is the area of the disk ending on the brane. Since we are considering a single brane therefore $U=e^{i\theta}$ and the sum over the partitions is restricted to partitions of type $\{(k)\,|\,k=0,1,2,\cdots\}$. The partition function is then given by
\bea
Z_{Brane}&=&\sum_{k=0}^{\infty}(-Q\,U)^{k}s_{(k)}(q^{-\rho})=\sum_{k=0}^{\infty}(-Q\,U)^{k}q^{k/2}\prod_{i=1}^{k}(1-q^{i})^{-1}\\\nn
&=&\prod_{n=1}^{\infty}\Big(1-Q\,U\,q^{n-\frac{1}{2}}\Big)\,.
\eea
Define $z=Q\,U\,q^{-\frac{1}{2}}$.  Then
\bea\nn
\Big|Z_{Brane}\Big|^{2}&=&\prod_{r=0}^{\infty}\Big(\frac{1-z\,q^{r+1}}{1-\bar{z}\,q^{r}}\Big)\,,
\eea
this is precisely the result given in \cite{DGG} (Eq.(3.6)) if we take $z=q^{\frac{m}{2}}\,\zeta$
for the generalized index \cite{Kapustin}  where $m$ is the monopole charge.

\subsubsection{Lagrangian brane on local $\mathbb{P}^{1}\times \mathbb{P}^{1}$}
Here we will consider a single Lagrangian brane on $\mathbb{P}^{1}\times \mathbb{P}^{1}$. The brane configuration is shown in \figref{lag1} below.

\begin{figure}[h]
  \centering
  \includegraphics[width=3.5in]{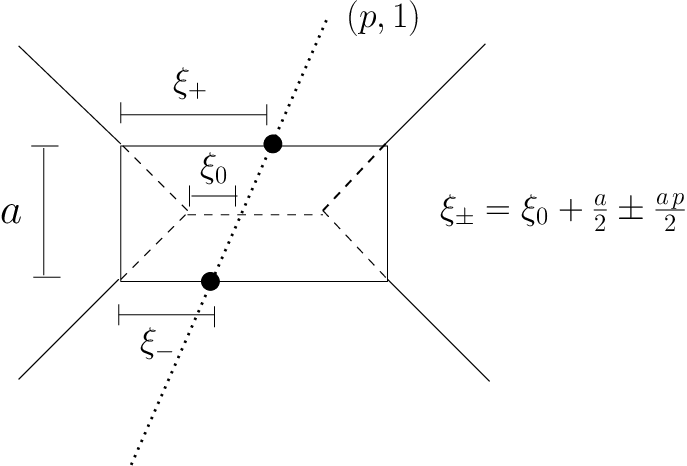}\\
  \caption{The geometry of Lagrangian brane on local $\mathbb{P}^{1}\times \mathbb{P}^{1}$.}\label{lag1}
\end{figure}

\begin{figure}[h]
  \centering
  \includegraphics[width=3.5in]{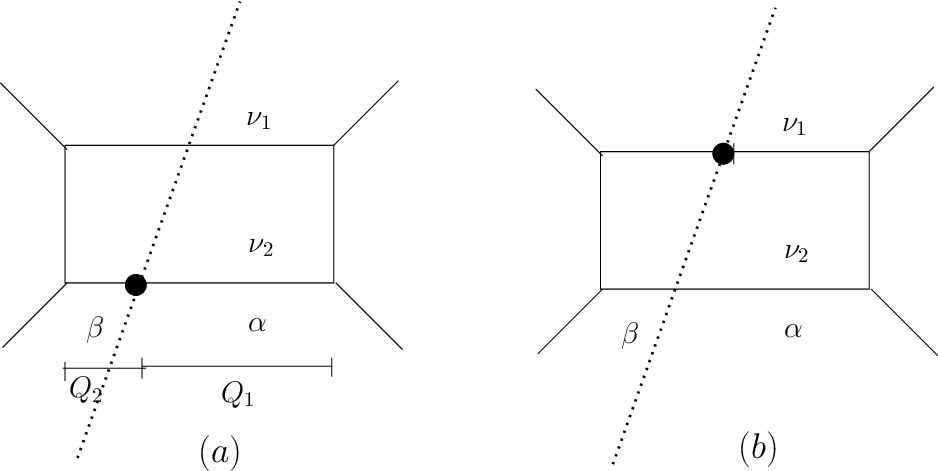}\\
  \caption{}\label{p1p1}
\end{figure}

In the limit $a\mapsto 0$ the 4-cycle collapses to the curve $B$ and the position of the Lagrangian brane on the $B$ is determined by $\xi_{0}$. When we deform away from this point there are two possibilities for the Lagrangian brane. Either it ends on the upper horizontal line or the lower one as shown in \figref{p1p1}. We will consider both possibilities in calculating the index.

The partition function of the brane depends on $(Q_{1},Q_{b},Q_{f})$, which in turn depend on the parameters of the geometry for a given fixed position of the external legs, and the holonomy on the brane . For the brane attached to the lower horizontal leg (see \figref{p1p1})
The
\bea
Q_{b}=uQ_{f}\,,\,\,\,\,\,\,\,Q_{1}=u\,e^{\xi_{0}}\,Q_{f}^{\frac{p+1}{2}}
\eea
The partition function of the geometry when the Lagrangian brane is on the upper horizontal leg is given by
\bea
Z^{p}_{Brane}(Q_{1}',Q_{b},Q_{f})
\eea
where
\bea
Q_{b}=u\,Q_{f}\\\nn
Q_{1}'=e^{-\xi_{0}}Q_{f}^{\frac{p+1}{2}}
\eea

To introduce a Lagrangian brane on one of the internal legs of this CY threefold we generalize the unrefined formalism of \cite{kanno}. Two new partitions $\alpha$ and $\beta$ are introduced to account for the new open strings. The open string partition function is given by
\bea\nn
Z_{Brane}&:=&\sum_{\nu_{1}\,\nu_{2}}(-Q_{b})^{|\nu_{1}|+|\nu_{2}|}(-Q_{1})^{|\alpha|}(-Q_{2})^{|\beta|}\,(f_{\nu_{1}\otimes \beta}(t,q))^{p-1}\,(f_{\nu_{1}^t\otimes \alpha}(q,t))^{p}\,f_{\nu_{2}^t}(t,q)\,\\\nn
&&P_{\nu_{1}\otimes \alpha}(t^{\rho};q,t)P_{\nu_{1}^t\otimes \beta}(q^{\rho};t,q)
P_{\nu_{2}}(q^{\rho};t,q)\,P_{\nu_{2}^t}(t^{\rho};q,t)\,
\mbox{Tr}_{\alpha}U\,\mbox{Tr}_{\beta}U^{-1}\\\nn
&&\prod_{i,j}\Big[\Big(1-Q_{f}\,t^{-i+1+\nu_{2,j}}\,q^{-j+(\nu_{1}\otimes \alpha)_{1,i}}\Big)\Big(1-Q_{f}q^{-i+1+(\nu_{1}\otimes \beta^t)_{1,j}}t^{-j+\nu_{2,i}}\Big)\Big]^{-1}\,,\eea
where $p$ is the framing of the brane. Since we are considering a single brane therefore the sum over $\alpha$ and $\beta$ is restricted to partitions of type $\{(n)\,|\,n=0,1,\cdots\}$. It is clear from \figref{p1p1} that
\bea
Q_{1}Q_{2}=Q_{b}\,.
\eea
therefore
\bea\nn
Z_{Brane}&:=&\sum_{\nu_{1}\,\nu_{2}}(-Q_{b})^{|\nu_{1}|+|\nu_{2}|+|\beta|}(-Q_{1})^{|\alpha|-|\beta|}\,(f_{\nu_{1}\otimes \beta}(t,q))^{p-1}\,(f_{\nu_{1}^t\otimes \alpha}(q,t))^{p}\,f_{\nu_{2}^t}(t,q)\,\\\nn
&&P_{\nu_{1}\otimes \alpha}(t^{\rho};q,t)P_{\nu_{1}^t\otimes \beta}(q^{\rho};t,q)
P_{\nu_{2}}(q^{\rho};t,q)\,P_{\nu_{2}^t}(t^{\rho};q,t)\,\mbox{Tr}_{\alpha}U\,\mbox{Tr}_{\beta}U^{-1}\\\nn
&&\prod_{i,j}\Big[\Big(1-Q_{f}\,t^{-i+1+\nu_{2,j}}\,q^{-j+(\nu_{1}\otimes \alpha)_{1,i}}\Big)\Big(1-Q_{f}q^{-i+1+(\nu_{1}\otimes \beta^t)_{1,j}}t^{-j+\nu_{2,i}}\Big)\Big]^{-1}\,.\,
\eea
Taking into account contributions of order $Q_{b}$ and $Q_{1}$ we get
\bea
Z^{p}_{Brane}(Q_{1},Q_{b},Q_{f})=Z_{0}\Big(\tilde{Z}-Q_{b}\Big(Z_{1}+Z_{2}+Z_{3}\Big)+\cdots\Big)\,.
\eea
Where
\bea
Z_{0}&=&\Big[\prod_{i,j}\Big[\Big(1-Q_{f}\,t^{i}\,q^{j-1}\Big)\Big(1-Q_{f}q^{i}\,t^{j-1}\Big)\Big]^{-1}\Big]\\\nn
\widetilde{Z}&:=&\sum_{\alpha}\,\Big(-Q_{1}\,U\Big)^{|\alpha|}(f_{\alpha}(q,t))^{p}\,P_{\alpha}(t^{\rho};q,t)\prod_{(i,j)\in \alpha}\Big(1-Q_{f}\,q^{-i}\,t^{j}\Big)^{-1}\\\nn
&=&\Big(1+\frac{Q_{1}UQ_{f}^{-1}\,q^{\frac{p}{2}+1}\,t^{-\frac{p}{2}-\frac{3}{2}}}{(1-t^{-1})(1-q\,t^{-1}\,Q_{f}^{-1})}-\\\nn
&&\frac{Q_{1}^{2}U^{2}Q_{f}^{-2}\,q^{p+3}\,t^{-2p-3}}{(1-t^{-1})(1-t\,q)(1-q\,t^{-1}\,Q_{f}^{-1})(1-q\,t^{-2}\,Q_{f}^{-1})}+\cdots\Big)\,,
\eea
and
\bea\nn
Z_{1}&=&-\frac{\frac{q}{t}}{(1-q)(1-t^{-1})(1-q\,t^{-1}\,Q_{f})}\times \\\nn
&&\sum_{\alpha}(-Q_{1}U)^{|\alpha|}\,(f_{\alpha}(q,t))^{p}\,P_{\alpha}(t^{\rho};q,t)
\prod_{i,j}\frac{1-Q_{f}t^{-i+1}q^{-j}}{1-Q_{f}t^{-i+1}q^{-j+(\Box\otimes \alpha)_{i}}}\\\nn
&=&\frac{q\,t^{-1}}{(1-q)(1-t^{-1})(1-Q_{f})(1-q\,t^{-1}\,Q_{f})}
\Big[1-\\\nn
&&Q_{1}U\frac{q^{\frac{p}{2}}\,t^{-\frac{p+1}{2}}}{(1-t^{-1})(1-Q_{f})(1-q\,Q_{f})(1-t^{-1}\,Q_{f})}\\\nn
&&-
Q_{1}^{2}\,U^{2}\,\frac{q^{p+1}t^{-2p}}{(1-t^{-1})(1-q\,t)(1-Q_{f})(1-t^{-1}\,Q_{f})(1-q\,Q_{f})^{2}(1-q^{2}\,Q_{f})}+\cdots\Big]\\\nn
\eea
\bea\nn
Z_{2}&=&\frac{1}{(1-q)(1-t^{-1})(1-Q_{f})}\times \\\nn
&&\sum_{\alpha}(-Q_{1}U)^{|\alpha|}\,(f_{\alpha}(q,t))^{p}\,P_{\alpha}(t^{\rho};q,t)
\prod_{i,j}\frac{1-Q_{f}t^{-i+1}q^{-j}}{1-Q_{f}t^{-i+1+\Box_{j}}q^{-j+\alpha_{i}}}\\\nn
&=&\frac{q\,t^{-1}}{(1-q)(1-t^{-1})(1-Q_{f})}\Big[-\frac{Q_{f}^{-1}}{(1-q\,t^{-1}\,Q_{f}^{-1})}-Q_{1}U\frac{q^{\frac{p}{2}-1}\,t^{-\frac{p-1}{2}}}{(1-t^{-1})(1-Q_{f})}\\\nn
&&-Q_{1}^{2}\,U^{2}\,\frac{q^{p}\,t^{-2p+1}}{(1-t^{-1})(1-q\,t)(1-Q_{f})(1-q\,Q_{f})}+\cdots\Big]\\\nn
\eea
\bea
Z_{3}&=&-\frac{Q_{1}^{-1}\,U^{-1}\,\Big(\sqrt{\frac{t}{q}}\Big)^{p-1}\sqrt{q}}{(1-q)(1-\frac{q}{t}Q_{f})}\sum_{\alpha}(-Q_{1}U)^{|\alpha|}\,(f_{\alpha}(q,t))^{p}\,
P_{\alpha}(t^{\rho};q,t)
\prod_{i=1}^{\alpha}(1-Q_{f}q^{\alpha-i})^{-1}\\\nn
&=&-\frac{Q_{1}^{-1}\,U^{-1}\,\Big(\sqrt{\frac{t}{q}}\Big)^{p-1}\sqrt{q}}{(1-q)(1-\frac{q}{t}Q_{f})}\Big[1-Q_{1}\,U\,\frac{q^{\frac{p}{2}}\,t^{-\frac{p+1}{2}}}{(1-Q_{f})}
-\\\nn
&&Q_{1}^{2}\,U^{2}\,\frac{q^{p+1}t^{-2p}}{(1-t^{-1})(1-q\,t)(1-Q_{f})(1-q\,Q_{f})}+\cdots\Big]
\eea
In the above equations $P_{\alpha}({\bf x};q,t)$ are Macdonald polynomials and since $\alpha$ only takes the values $\{(m)\,|\,m=0,1,\cdots\}$ we give below the explicit expression for $P_{(m)}(t^{\rho};q,t)$ as a function of $x$ and $y$ which we will need later:
\bea
P_{(m)}(t^{\rho};q,t)&=&(-1)^{m}t^{m/2}q^{m(m-1)/2}\prod_{j=1}^{m}\Big(1-t\,q^{m-j}\Big)^{-1}\\\nn
&=&(-1)^{m-1}\,\frac{y^{\frac{m^2}{2}-1}\,x^{\frac{m^{2}}{2}+1-m}}{(1-x\,y^{-1})\prod_{j=1}^{m-1}\Big(1-y^{m-j+1}x^{m-j-1}\Big)}\,\\\nn
P_{(m)}(q^{\rho};t,q)&=&-\frac{x^{1-\frac{m}{2}}\,y^{1-\frac{m}{2}}}{(1-x\,y)\prod_{j=1}^{m-1}(1-x^{m-j-1}y^{j-m-1})}
\eea

The partition function of the geometry with Lagrangian brane has a closed string factor which is the partition function of the geometry without the brane and an open string factor,
\bea
Z_{Brane}=Z_{closed}\times Z_{open}
\eea
In our case there are two different possibilities for the brane to end when the loop variable is deformed. We denote the open string partition function of the brane on the lower leg by $Z_{open}$ and the open string partition function of the brane on the upper leg by $\widetilde{Z}_{open}$. The two are related as follows:
\bea
\widetilde{Z}_{open}=Z_{open}(e^{-\xi_{0}}Q_{f}^{\frac{p+1}{2}}U^{-1}, u\,Q_{f},Q_{f})\,,
\eea
where the open string partition function of the brane on the lower leg is
\bea\nn
Z_{open}(u\,e^{\xi_{0}}\,Q_{f}^{\frac{p+1}{2}}\,U, u\,Q_{f},Q_{f})\,.
\eea
The index of the defect theory is given by
\bea
I_{p}&=&\int \frac{1}{2}\frac{dQ_{f}}{2\pi iQ_{f}}\frac{dU}{2\pi i U}\,\Big|\,Z_{open}\,\widetilde{Z}_{open}\Big|^2\\\nn
&=&\int \frac{1}{2}\frac{dQ_{f}}{2\pi iQ_{f}}\frac{dU}{2\pi i U}\,\Big|Z_{open}(u\,e^{\xi_{0}}\,Q_{f}^{\frac{p+1}{2}}\,U, uQ_{f},Q_{f})\,Z_{open}(e^{-\xi_{0}}Q_{f}^{\frac{p+1}{2}}U^{-1}, uQ_{f},Q_{f})\Big|^2\\\nn
\eea
where under complex conjugation
\bea
(e^{-\xi_{0}},u,Q_{f},U)\mapsto (e^{-\xi_{0}},u^{-1},Q_{f}^{-1},U^{-1})\,.
\eea
For $p=1$ the above index up to order $x^2$ is $(v=e^{\xi_{0}})$
\bea\nn
I_{p=1}&=&\Big[1+\frac{x}{u^3\,y}+\Big(\frac{2}{u^3}+\frac{4}{u^2}-\frac{1}{u\,v^2}-\frac{v^2}{u}
+\frac{1}{u^6\,y^2}+\frac{4}{u^3\,y^2}-
\frac{10y^2}{u^4\,(1-y^2)^2}
-\\\nn
&&\frac{12}{u^5\,(1-y^2)}
-\frac{2}{u\,(1-y^2)}+\frac{2y^2}{u\,(1-y^2)}\Big)x^2+\cdots\Big]
\eea
To compute the index in monopole sector $m$ we simply substitute $v=x^{m/2}$
\section{Conclusion}

We have seen in this paper that the BPS states which arise in the IR
flow of superconformal theories upon deformations, are a powerful tool
in computing superconformal indices at the conformal point.  This is particularly
so in $d=3,5$, where we have proposed how one may recover the full index
in terms of the BPS partition functions.  Even though we have not given a full
derivation of the proposal we have checked that it works in all the known examples.
It should be possible\footnote{We conjecture that this should make sense in the full M-theory
context, at least for non-compact Calabi-Yau, i.e., that one could embed the constructions
of \cite{pestun,nati} and similar extensions in other dimensions in the full string theory.} to derive these results by compactifying M-theory on
toric 3-folds times $S^1\times S^4$ and
applying localization ideas to the full string theory similar to the derivation of OSV conjecture
in \cite{beas}.

It is natural to ask whether we can compute the partition function of supersymmetric
theories on $S^5$ and $S^3$ using BPS data.  Indeed there is a natural proposal
for this \cite{guli}, which shows how this may be done using topological strings.  Moreover this can also be used to formulate the index of $(1,0)$ and $(2,0)$ theories on $S^1\times S^5$.
We thus see that BPS states, as captured by topological strings, are powerful enough to
capture the partition function and the superconformal index of a large number of
theories in diverse dimensions.

Our work gives further motivation for a reformulation of supersymmetric theories
entirely in terms of their BPS data in the IR, in diverse dimensions with varying amounts
of supersymmetry.  It would be very important to see if one can fully reconstruct
the superconformal theories solely from their BPS data.

\section*{Acknowledgments}

We would like to thank M. Aganagic, S. Cecotti, C. Cordova, T. Dimofte, A.Gadde, S. Gukov, J. Heckman, Y. Imamura,
K. Intriligator, D. Jefferis, G. Lockhart, S. Minwalla, V. Pestun, L. Rastelli and N. Seiberg for useful discussions.
C.V. would also like to thank the Simons Center for Geometry and Physics where he attended the 10-th Simons Workshop on math and physics.
The work of A.I. is  supported in part by the Higher Education Commission grant HEC-2052.
The work of C.V. is supported in part by NSF grant PHY-0244821.

\bibliography{physics}

\end{document}